\documentclass[aps,twocolumn,showpacs,preprintnumbers,amsmath,amssymb]{revtex4}

\usepackage{graphicx}
\usepackage{dcolumn}
\usepackage{bm}
\usepackage{txfonts}
\usepackage{lscape}
\begin{document}

\title{Functional renormalization for spontaneous symmetry breaking in the Hubbard model}

\author{S. Friederich}  
  \thanks{new address: Universit\"at Wuppertal, Fachbereich C -- Mathematik und Naturwissenschaften, Gau\ss str. 20, D-42119 Wuppertal, Germany}
\author{H. C. Krahl}  
\author{C. Wetterich}

\affiliation{\mbox{\it Institut f{\"u}r Theoretische Physik,
Universit\"at Heidelberg,
Philosophenweg 16, D-69120 Heidelberg, Germany}}

\begin{abstract}
The phases with spontaneously broken symmetries corresponding to antiferromagnetic and $d$-wave superconducting order in the two-dimensional $t-t'$-Hubbard model are investigated by means of the functional renormalization group. The introduction of composite boson fields in the magnetic, charge density and superconducting channels allows an efficient parametrization of the four-fermion vertex and the study of regimes where either the antiferromagnetic or superconducting order parameter, or both, are nonzero. We compute the phase diagram and the temperature dependence of the order parameter below the critical temperature, where antiferromagnetic and superconductiving order show a tendency of mutual exclusion.
\end{abstract}


\maketitle

\section{Introduction}

The two-dimensional Hubbard model \cite{hubbard,kanamori,gutzwiller} on a square lattice is widely believed to hold a key role for the understanding of high-temperature superconductivity in the cuprates. In analogy to the cuprate phase diagram it seems to exhibit antiferromagnetic and $d$-wave superconducting order in close vicinity \cite{anderson,miyake,loh,bickers,lee,millis,monthoux,scalapino,bickersscalapinowhite,bulut,pruschke,grote,hankevych,maier,senechal,maierjarrellscalapino,maiermacridin}; for a systematic overview, see Ref. \onlinecite{scalapinoreview}. The tendency toward $d$-wave superconductivity was already predicted by some strikingly simple scaling approaches \cite{schulz,dzyaloshinskii,lederer}. On a higher level of technical sophistication, the fermionic functional renormalization group approach \cite{zanchi1,zanchi2,halbothmetzner,halbothmetzner2,salmhoferhonerkamp,honerkamp01,honisalmi,katanin} has been of great help to analyze in detail the competition of different types of instabilities.

The main result of the present paper concerns the phase diagram in the low temperature regime, where spontaneous symmetry breaking occurs either in the antiferromagnetic or in the $d$-wave superconducting channel. The symmetry broken regimes are difficult to access for fermionic functional renormalization group studies such as  \cite{zanchi1,zanchi2,halbothmetzner,halbothmetzner2,salmhoferhonerkamp,honerkamp01,honisalmi,katanin}. Renormalized mean field investigations \cite{metznerreissrohe,reiss}, based on a combination of the fermionic functional renormalization group approach with a mean field treatment, have already been able to study the mutual influence of the order parameters for antiferromagnetism and $d$-wave superconductivity. The functional renormalization group approach presented in this paper makes it possible to investigate this problem by entering the spontaneously broken regimes in a partially bosonized language where the different types of bosonic fields that are introduced correspond to the different types of possible order of the system.

\begin{figure}[t]
\includegraphics[width=85mm,angle=0.]{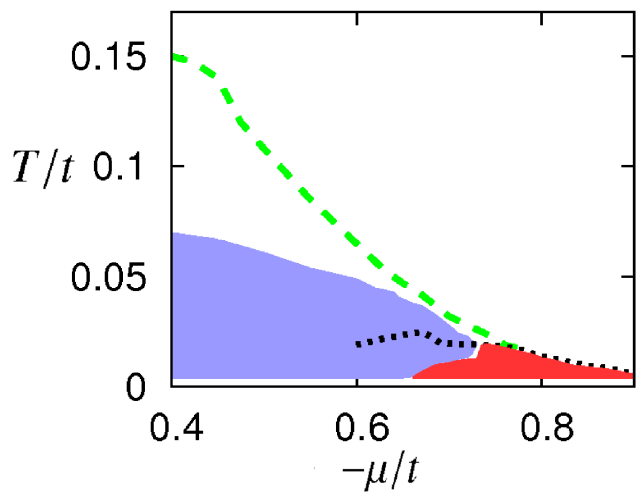}
\caption{\small{Phase diagram for $U/t=3$ and $t'/t=-0.1$. The (blue) solid region for smaller values of $|\mu|$ denotes the antiferromagnetic phase, the (red) solid region for larger values of $|\mu|$ the $d$-wave superconducting phase. The green dashed line and the black dotted line indicate the pseudocritical temperatures below which local order sets in for antiferromagnetism and $d$-wave superconductivity, respectively. The pseudocritical line for antiferromagnetism ends at $-\mu/t\approx0.79$. The region below $T_{min}=4\cdot10^{-3}t$ has been left blank since calculations were only done for higher temperatures. The range of the chemical potential $\mu$ shown here corresponds to electron filling between $0.9$ (for $-\mu/t\approx0.4$) and $0.7$ (for $-\mu/t\approx0.9$) per site. (We have no precise estimate for $\langle n_e\rangle$ as a functiona of $\mu$ for all $T$.)}}
\label{phasediag}
\end{figure}

We find a region of electron fillings with competition between antiferromagnetic and superconducting order. The two phenomena show a strong tendency of mutual exclusion. Nevertheless, we find regions where the two types of \textit{local} order coexist, even though this coexistence may not be maintained on the length scales of global order. We also find near the van Hove filling a considerable range of temperatures with local but not global antiferromagnetic order. For smaller electron filling the critical temperatures for the onset of local and global superconducting order almost coincide. An overview of our findings in form of the $\mu-T$-phase diagram (for next-to-nearest neighbor hopping $t'/t=-0.1$ and Hubbard repulsion $U/t=3$) is shown in Fig. \ref{phasediag}.

The approach presented in this work focuses on the low temperature behavior and brings together earlier attempts \cite{bbw00,bbw04,bbw05,kw07,krahlmuellerwetterich,simon,fourpoint} to perform a functional renormalization group analysis of the two-dimensional Hubbard model based on partial bosonization (or Hubbard-Stratonovich transformation) \cite{hubbardtransf,stratonovich}. In analogy to the parametrization methods for the fermionic four-point vertex proposed and developed in \cite{husemann,eberlein}, this method makes it possible to treat the complex momentum dependence of this function in an efficient, simplified way, involving only a comparatively small number of coupled flow equations. The fermionic four-point vertex, which is a scale-dependent function of three independent momenta, is decomposed in terms of bosonic propagators and Yukawa couplings, which are each functions of only one variable.

The main advantage of the method used in the present work consists of the possibility of following the renormalization group flow into regimes where one or more symmetries of the Hubbard Hamiltonian are spontaneously broken. (For other renormalization group studies of symmetry broken phases in similar models see \cite{honisalmilauschi,ossadnik,strack}.) At a certain scale $k_{SSB}$ of the renormalization flow, the momentum-dependent fermionic four-point vertex may diverge, and this signals the onset of local collective order. In order to extend the renormalization group treatment to the locally ordered regimes, composite degrees of freedom such as magnons and Cooper pairs are made explicit in terms of composite bosons. These correspond to different types of collective order the system might exhibit. A nonzero expectation value of the magnon field, for instance, signals the presence of some form of magnetic order, and a nonzero value of some Cooper pair field signals superconducting order. Different Cooper pair fields are distinct due to different symmetries of the order parameters they correspond to. The language of partial bosonization, where the different types of bosons are taken into account explicitly, is therefore the right tool to investigate the regimes exhibiting different forms of collective order. A particular advantage of the present approach, which combines functional renormalization and partial bosonization, is that it allows to investigate the possible coexistence of different types of order in the same range of parameters.

Earlier renormalization group studies using the framework of partial bosonization (see Refs. \onlinecite{bbw04, bbw05, krahlmuellerwetterich, simon}) incorporate only a comparatively small number of bosonic fields and therefore obtain a poorer resolution of the four-fermion vertex. In addition, they suffer from the so-called mean-field ambiguity \cite{jaeckelw03} which refers to the fact that by the use of the Fierz relations the microscopic Hubbard action can be mapped by different Hubbard-Stratonovich transformations into various equivalent descriptions involving different bosonic fields. In the presence of approximations the final results eventually depend on the specific choice of the decomposition. As argued in Ref. \onlinecite{fourpoint}, this shortcoming can be avoided in the present approach by explicitly keeping the original Hubbard repulsion $U$ in the truncation. Contributions to the four-fermion vertex that emerge during the flow are attributed to the different bosonic channels, depending on their momentum structure and therefore in an essentially unbiased way. A remaining bias concerns the choice of bosons taken into account. We choose to include four different bosons corresponding to magnetic, charge density, $s$-wave and $d$-wave superconducting order. This choice can be motivated by the structure of the one-particle-irreducible (1PI) diagrams from which the contributions to the flow of the fermionic four-point vertex are derived \cite{fourpoint}.

Our truncation for the flowing action includes, for the symmetric regime, the parametrization of the four-fermion vertex already used in Ref. \onlinecite{fourpoint}. Furthermore it contains quartic bosonic couplings for the antiferromagnetic and $d$-wave superconducting bosons as well as self-energy corrections for the fermions in terms of a frequency-dependent wave function renormalization. We observe symmetry breaking within the studied parameter region only in the antiferromagnetic and $d$-wave superconducting channels. In these symmetry broken regimes, the $\rho$- and $s$-bosons are dropped from the truncation for renormalization scales $k<k_{SSB}$, and we focus on the fermionic and bosonic contributions to the order parameters and quartic couplings of the antiferromagnetic and $d$-wave superconducting bosons.

\section{Method and approximation}
The starting point of our treatment is the exact flow equation for the effective average action or flowing action \cite{cw93}:
\begin{equation} \label{floweq}
\partial_k \Gamma_k = \frac{1}{2} \rm{STr} \,
  \left(\Gamma^{(2)}_k + R_k\right)^{-1}  \partial_k R_k =
 \frac{1}{2} \rm{STr}\,\tilde \partial_k \,\ln  (\Gamma^{(2)}_k + R_k)\,.
\end{equation}
The dependence on the renormalization scale $k$ is introduced by adding a regulator $R_k$ to the full inverse propagator $\Gamma^{(2)}_k$. In Eq. (\ref{floweq}) $\rm{STr}$ denotes a supertrace, which sums over momenta, frequencies, and internal indices, containing a minus sign for fermions, while $\tilde \partial_k=(\partial_kR_k)\frac{\partial}{\partial R_k}$ is the scale derivative acting only on the scale dependence introduced by the regulator $R_k$. The Hamiltonian of the system under considerations is taken into account by the initial condition $\Gamma_{k=\Lambda}=S$ of the renormalization flow, where $\Lambda$ denotes some very large UV scale and $S$ is the microscopic action in a functional integral formulation of the Hubbard model. For the Hubbard model, this action is given by
\begin{eqnarray}\label{Hubbardaction}
S&=&\sum_Q\hat\psi^{\dagger}(Q)[i\omega_Q+\xi_Q]\hat\psi(Q)\\
&&+\frac{U}{2}\sum_{K_1,K_2,K_3,K_4}\big\lbrack\hat\psi^\dagger(K_1)\hat\psi(K_2)\big\rbrack\,\big\lbrack\hat\psi^\dagger(K_3)\hat\psi(K_4)\big\rbrack\nonumber\\
&&\hspace{2cm}\times\delta\left( K_1-K_2+K_3-K_4 \right)\,,\nonumber
\end{eqnarray}
where
\begin{equation}
\hat\psi(Q)=\left(\hat\psi_\uparrow(Q),\\\hat\psi_\downarrow(Q) \right)^T
\end{equation}
are Grassmann fields describing electrons on a square lattice. The next-neighbor and next-to-nearest-neighbor hopping parameters $t$ and $t'$ are reflected in
\begin{equation}
\xi(\mathbf q)=-\mu-2t(\cos q_x +\cos q_y)-4t' \cos q_x\cos q_y\,.
\end{equation}

Here, as well as in all that follows, we employ a compact notation with $Q=(\omega_n=2\pi nT,\mathbf{q})$ for bosonic and $Q=(\omega_n=(2n+1)\pi T,\mathbf{q})$ for fermionic fields, and
\begin{eqnarray}\label{eq:sumdefinition}
&&\quad\sum\limits_Q=T\sum\limits_{n=-\infty}^\infty \int\limits_{-\pi}^\pi \frac{d^2q}{(2\pi)^2}\,,\nonumber\\
&&\delta(Q-Q')=T^{-1}\delta_{n,n'}(2\pi)^2\delta^{(2)}(\mathbf{q}-\mathbf{q'})\,.
\end{eqnarray}
The components of the momentum $\mathbf q$ are measured in units of the inverse lattice distance $\mathrm{a}^{-1}$. The discreteness of the square lattice is reflected by the $2\pi$-periodicity of the momenta $\mathbf{q}$.

In the limit $k\to 0$ all fluctuations are included and the flowing action $\Gamma_k$ equals the full effective action $\Gamma=\Gamma_{k\to 0}$, which is the generating functional of the one-particle irreducible (1PI) vertex functions. For $k>0$ the bosonic fluctuations with momenta $|\mathbf q|<k$ and the fermionic fluctuations with momenta $|\mathbf q-\mathbf q_F|<k$, where $\mathbf q_F$ is the Fermi-momentum vector which is closest to $\mathbf q$, are not yet included.

Although Eq. \eqref{floweq} is an exact flow equation, it can only be solved approximately. In particular, a truncation has to be specified for the flowing action, indicating which of the (infinitely many) 1PI vertex functions are actually taken into account. We use different truncations for the disordered symmetric regime (SYM) and the spontaneously broken regimes (SSB) where one of the bosonic fields has a nonzero expectation value. In what follows we denote the regime where both the order parameter $\alpha_0$ for antiferromagnetism and $\delta_0$ for $d$-wave superconductivity are nonzero by SSBad. The regimes where only either $\alpha_0$ or $\delta_0$ is nonzero, the other one being zero, are denoted by SSBa and SSBd, respectively. The nonzero expectation values $\alpha_0$ or $\delta_0$ in the SSB regimes indicate local order.Since we are dealing with a two-dimensional model, the order parameters $\alpha_0$ and $\delta_0$ must become zero in the thermodynamic limit for $k\mapsto0$, in accordance with the Mermin-Wagner theorem. They may, however, remain nonzero for $k<l^{-1}$, with $l$ the size of a typical experimental probe, signaling the appearance of long-range, that is ``global'', (for instance magnetic or superconducting) order on macroscopic length scales.

Our ansatz for the flowing action includes terms for the electrons, for the bosons in the magnetic, charge, $s$-wave and $d$-wave superconducting channels, and for interactions between fermions and bosons:
\begin{eqnarray}
\Gamma_k[\chi]&=&\Gamma_{F,k}+\Gamma_{Fm,k}+\Gamma_{F\rho,k}+\Gamma_{Fs,k}+\Gamma_{Fd,k}\\
       &&+\Gamma_{a,k}+\Gamma_{\rho,k}+\Gamma_{s,k}+\Gamma_{d,k}+\sum_XU_{B,k}(\mathbf a,\rho,s,d)\,.\nonumber
\end{eqnarray}
The collective field $\chi=(\mathbf a,\rho,s,s^*,d,d^*,\psi,\psi^*)$ includes both fermion fields $\psi,\psi^*$ and boson fields $\mathbf a,\rho,s,s^*,d,d^*$. The ``antiferromagnetic'' boson field $\mathbf a(Q)$ is related to the ``magnetic'' boson field $\mathbf m(Q)$ used in Ref. \onlinecite{fourpoint} by a shift in the momentum variable with respect to the antiferromagnetic wave vector $\Pi=(0,\pi,\pi)$,
\begin{eqnarray}
\mathbf a(Q)=\mathbf m(Q+\Pi)\,.
\end{eqnarray}

The purely fermionic part $\Gamma_{F}$ (the dependence on the scale $k$ is always implicit in what follows) of the flowing action consists of a fermion kinetic term $\Gamma_{F\rm{kin}}$, a momentum-independent four-fermion term $\Gamma_F^U$, and the momentum-dependent four-fermion terms $\Gamma_F^a,\Gamma_F^\rho,\Gamma_F^s,\Gamma_F^d$:
\begin{eqnarray}
\Gamma_{F}=\Gamma_{F\rm{kin}}+\Gamma_{F}^U+\Gamma_F^a+\Gamma_F^\rho+\Gamma_F^s+\Gamma_F^d\,.
\end{eqnarray}
The fermionic kinetic term is essentially left unchanged with respect to the original Hubbard Hamiltonian \eqref{Hubbardaction}, apart from the fact that a fermionic wave function renormalization is included, which depends on the Matsubara frequency,
\begin{eqnarray}\label{fermprop}
\Gamma_{F\rm{kin}}=\sum_{Q}\psi^{\dagger}(Q)P_F(Q)\psi(Q)\,,
\end{eqnarray}
where
\begin{eqnarray}\label{PF}
P_{F}(Q)=Z_F(\omega_Q)\left(i\omega_Q+\xi(\mathbf q)\right)
\,.\end{eqnarray}
On initial scale $k=\Lambda$ we set $Z_F(\omega_Q)=1$ for all frequencies in order to equal the kinetic term of the microscopic Hubbard action, see Eq. \eqref{Hubbardaction}. The flow of $Z_F(\omega_Q)$ is neglected for all frequencies except for the two lowest Matsubara modes $\omega_Q=\pm\pi T$. The computation of the scale-dependent quantity $Z_F(\pm\pi T)$ is described in the following section. Self-energy corrections to the dependence of $P_{F}(Q)$ on spatial momentum are omitted. According to \cite{giering} the influence of self-energy corrections due to the frequency dependence of $P_{F}(Q)$ seems to be more important.

As motivated  in the introduction, the momentum-independent part of the four-fermion coupling, which at $k=\Lambda$ is identical to the Hubbard interaction $U$, remains unmodified during the flow. The corresponding part of the effective action reads in our truncation
\begin{eqnarray}
\Gamma_{F}^U &=&\frac{1}{2}\sum_{K_1,K_2,K_3,K_4}U\,\delta\left( K_1-K_2+K_3-K_4 \right)\,\nonumber\\&&\hspace{0.5cm}\times\,\big\lbrack\psi^\dagger(K_1)\psi(K_2)\big\rbrack\,\big\lbrack\psi^\dagger(K_3)\psi(K_4)\big\rbrack\,.
\end{eqnarray}

We focus on the momentum and spin dependence of the fermionic four-point function $\lambda_F(K_1,K_2,K_3,K_4)$, which, due to energy-momentum conservation, is a function of three independent momenta (e.g., $K_4=K_1-K_2+K_3$). We decompose this vertex into a sum of four functions $\lambda_F^a(Q)$, $\lambda_F^\rho(Q)$, $\lambda_F^s(Q)$ and $\lambda_F^d(Q)$, each depending on only one particular combination of the $K_i$. The chosen decomposition of the fermionic four-point function is inspired by the singular frequency and momentum structure of the leading contributions during the renormalization flow, see Eqs. (9)-(12) in Ref. \onlinecite{fourpoint} for precise definitions. In our approach, these functions are described by the exchange of the four different bosons $\mathbf a$, $\rho$, $s$ and $d$.

Practically, this is achieved by the technique of flowing bosonization \cite{GiesWett,pawlowski,floerchi}, which was adapted to our purposes in Refs. \onlinecite{krahlmuellerwetterich,fourpoint}. The basic idea is to introduce scale-dependent bosonic fields in order to transform the momentum-dependent four-fermion vertex into Yukawa-type interactions between the fermions and bosons. In this way we keep the terms $\Gamma^a_F$, $\Gamma^\rho_F$, $\Gamma^s_F$ and $\Gamma^d_F$ at zero during the flow and replace their effects by flowing Yukawa interactions between fermions and bosons. These interaction terms read in our truncation
\begin{eqnarray}\label{GFak}
\Gamma_{Fa}&=&-\!\sum_{K,Q,Q'}\bar h_a(K)\;\mathbf{a}(K)\cdot[\psi^{\dagger}(Q)\boldsymbol{\sigma}\psi(Q')]\;\nonumber\\&&\hspace{2cm}\delta(K-Q+Q'+\Pi)\,,\nonumber\\
\Gamma_{F\rho}&=&-\!\sum_{K,Q,Q'}\bar h_\rho(K)\;\rho(K)\,[\psi^{\dagger}(Q)\psi(Q')]\;\delta(K-Q+Q')\,,\nonumber\\
\Gamma_{Fs}&=&-\!\sum_{K,Q,Q'}\bar h_s(K)\,\left(s^*(K)\,[\psi^{T}(Q)\epsilon\psi(Q')]\right.\\
&&\hspace{1.5cm}\left.-s(K)\,[\psi^{\dagger}(Q)\epsilon\psi^*(Q')]\right)\;\delta(K-Q-Q')\,,\nonumber\\
\Gamma_{Fd}&=&-\!\sum_{K,Q,Q'}\bar h_d(K)f_d\,\left((Q-Q')/2\right)\left(d^*(K)\,[\psi^{T}(Q)\epsilon\psi(Q')]\right.\nonumber\\
&&\hspace{1.5cm}\left.-d(K)\,[\psi^{\dagger}(Q)\epsilon\psi^*(Q')]\right)\;\delta(K-Q-Q')\,,\nonumber
\end{eqnarray}
where $\boldsymbol\sigma=(\sigma^1,\sigma^2,\sigma^3)^T$ is the vector of the Pauli matrices and $\epsilon=i\sigma^2$. The Yukawa-couplings $\bar h_a(Q)$, $\bar h_\rho(Q)$, $\bar h_s(Q)$, and $\bar h_d(Q)$ are running couplings which vanish on the initial scale $k=\Lambda$ of the renormalization flow since there is no non-trivial momentum dependence in the initial four-fermion term in the Hubbard action Eq. \eqref{Hubbardaction}. Note the presence of the d-wave form factor
\begin{eqnarray}f_d(Q)=f_d(\mathbf q)=\frac{1}{2}\left( \cos(q_x)-\cos(q_y) \right)\end{eqnarray}
in the second-to-last line of Eq. \eqref{GFak}, which is kept fixed on all scales.

The purely bosonic part of our truncation for the effective average action consists of the bosonic kinetic terms together with the bosonic effective potential. The kinetic terms of the bosons are defined as the momentum-dependent pieces $P_{i}(Q)$ of the inverse bosonic propagators. The inverse propagator of, for instance, the antiferromagnetic boson is given by $\tilde P_{a}(Q)\equiv P_{a}(Q)+\bar m_a^2$, where $\bar m_a^2$ is its minimal value and $P_{a}(Q)$ the (strictly positive) kinetic term which we parametrize as
\begin{eqnarray}
P_{a}(Q)=Z_a\omega_Q^2+A_aF(\mathbf q)\,.
\end{eqnarray}
In this equation we employ for $F(\mathbf q)$
\begin{eqnarray}\label{eq:apropparam1}
F_c(\mathbf q)=\frac{D_a^2\cdot[\mathbf{q}]^2}{D_a^2+[\mathbf{q}]^2}
\,,\end{eqnarray}
if commensurate antiferromagnetic fluctuations dominate. Here $[\mathbf{q}]^2$ is defined as $[\mathbf{q}]^2=q_x^2+q_y^2$ for $q_{x,y}\in [-\pi,\pi]$ and continued periodically otherwise. If incommensurate antiferromagnetic fluctuations dominate, we use
\begin{eqnarray}\label{eq:apropparam2}
F_i(\mathbf q,\hat q)&=&\frac{D_a^2\tilde F(\mathbf q,\hat q)}{D_a^2+\tilde F(\mathbf q,\hat q)}
\,,\end{eqnarray}
where the momentum dependence is quartic in momentum and explicitly includes the incommensurability $\hat q$:
\begin{eqnarray}\label{eq:apropparaminkomm}
\tilde F(\mathbf q,\hat q)=\frac{1}{4\hat q^2}\big((\hat q^2-[\mathbf q]^2)^2+4[q_x]^2[q_y]^2\big)
\,.\end{eqnarray}
The shape coefficient $D_a$ used in Eqs. \eqref{eq:apropparam1} and \eqref{eq:apropparam2} is defined as
\begin{eqnarray}
D_a=\frac{1}{A_a}\left(\tilde P_a(0,\pi,\pi)-\tilde P_a(0,\hat q,0)\right).
\end{eqnarray}
For the prescriptions used in the computation of $Z_a$ and $A_a$ in the symmetric regime and the parametrizations of the kinetic terms of the other bosons see Eqs. (B6)\,-\,(B8) in Ref. \onlinecite{fourpoint}. No linear frequency term is included in the kinetic terms of the superconducting bosons.

The contributions to the effective average action where the bosonic kinetic terms appear are
\begin{eqnarray}
\Gamma_{a}&=&\frac{1}{2}\sum_{Q}\mathbf{a}^{T}(-Q)P_a(Q)\mathbf{a}(Q)\,,\label{m_masse}\\
\Gamma_{\rho}&=&\frac{1}{2}\sum_{Q}\rho(-Q)P_\rho(Q)\rho(Q)\,,\\
\Gamma_{s}&=&\sum_{Q}s^*(Q)P_{s}(Q)s(Q)\,,\\
\Gamma_{d}&=&\sum_{Q}d^*(Q)P_{d}(Q)d(Q)\,.
\end{eqnarray}
One can reconstruct the momentum-dependent four-fermion interactions $\Gamma^i_F$ by solving the field equation for the bosons $i$ as a functional of fermionic variables (as derived by variation of $\Gamma$ with respect to the field for the boson $i$) and reinserting this functional into $\Gamma$. Our results for the fermionic four-point function are one-loop exact in the sense that the scale derivatives of all contributions up to second order in the Hubbard interaction $U$ are taken into account by our truncation, including their full dependence on spatial momentum.

We may summarize that in one-loop order the complicated spin and momentum dependence of the fermionic four-point function, as it emerges during the renormalization flow, is completely expressed by the bosonic propagators and Yukawa couplings connecting the fermions to the different bosons. We expect that also beyond one-loop order the dominant features of the momentum dependence of $\lambda_F$ are reasonably well reproduced by the solution of the flow equations in our truncation.

We also include in our truncation a local effective potential $U_B(\mathbf a,\rho,s,d)$ (not to be confused with the Hubbard interaction $U$). Here we make an expansion in powers of fields $\mathbf a$, $\rho$, $s$, and $d$ up to second order in $\rho$ and $s$ and up to the fourth order in $\mathbf a$ and $d$. This expansion has its limitations. For instance, it cannot describe first order transitions between two different phases with the same symmetries. For our purposes the polynomial expansion is expected to work reasonably well. Spontaneous symmetry breaking in the antiferromagnetic or superconducting channels can be described by a minimum of $U_B$ away from the origin in $\mathbf a$-$d$-space. In case of a second order phase transition this means that the terms quadratic in the fields $\mathbf a$ and $d$, $d^*$, evaluated at a macroscopic scale $k_{ph}$, turn negative for temperatures below the critical temperature $T<T_c$ and vanish for $T=T_c$. We denote the quartic coupling in the antiferromagnetic channel by $\bar\lambda_a$, the coupling in the $d$-wave superconducting channel by $\bar\lambda_d$, and the coupling between these two channels by $\bar\lambda_{ad}$. In the symmetric regime SYM we expand the effective potential around the zero value of the fields:
\begin{eqnarray}\label{Utrunc1}
\sum_XU_B(\mathbf a,\rho,s,d)&=&\sum_Q\frac{1}{2}\left(\bar m_a^2\,\mathbf{a}^{T}(-Q)\mathbf{a}(Q) + \bar m_\rho^2\,\rho(-Q)\rho(Q)\right)\nonumber \\
&&+ \bar m_s^2\,s^*(Q)s(Q) + \bar m_d^2\,d^*(Q)d(Q) \nonumber \\
&&+ \frac{1}{2}\sum_{Q_1,Q_2,Q_3,Q_4}\delta\left(Q_1+Q_2+Q_3+Q_4\right)\nonumber\\
&& \times\left(\bar\lambda_a\,\alpha(Q_1,Q_2)\alpha(Q_3,Q_4)\right.\\
&&+\bar\lambda_d\,\delta(Q_1,Q_2)\delta(Q_3,Q_4)\nonumber \\
&&\left.+2\bar\lambda_{ad}\,\alpha(Q_1,Q_2)\delta(Q_3,Q_4) \right)\,,\nonumber
\end{eqnarray}
where we have defined the quantities $\alpha(Q_1,Q_2)=\frac{1}{2}\mathbf a(Q_1)\cdot\mathbf a(Q_2)$ and $\delta(Q_1,Q_2)=d^*(Q_1)d(Q_2)$ (which has to be distinguished from the Dirac delta-function by the number of arguments).

In the spontaneously broken regime SSBad the minimum of the effective potential occurs at nonzero values of the fields $\mathbf a$ and $d$. In this case, we neglect the $\rho$- and $s$-bosons in our truncation and expand around the minimum of the effective potential at $(\alpha_0,\delta_0)$:
\begin{eqnarray}\label{Utrunc2}
\sum_XU_B(\mathbf a,d)&=& \frac{1}{2}\sum_{Q_1,Q_2,Q_3,Q_4}\delta\left(Q_1+Q_2+Q_3+Q_4\right)\nonumber\\
&&\Big(\bar\lambda_a\big\lbrace\alpha(Q_1,Q_2)-\alpha_0\delta(Q_1)\delta(Q_2)\big\rbrace\nonumber\\
&&\hspace{0.8cm}\times\big\lbrace\alpha(Q_3,Q_4)-\alpha_0\delta(Q_3)\delta(Q_4)\big\rbrace \nonumber \\
&&+\bar\lambda_d\big\lbrace\delta(Q_1,Q_2)-\delta_0\delta(Q_1)\delta(Q_2)\big\rbrace\\
&&\hspace{0.8cm}\times\big\lbrace\delta(Q_3,Q_4)-\delta_0\delta(Q_3)\delta(Q_4)\big\rbrace\nonumber\\
&&+2\bar\lambda_{ad}\big\lbrace\alpha(Q_1,Q_2)-\alpha_0\delta(Q_1)\delta(Q_2)\big\rbrace\nonumber\\
&&\hspace{0.8cm}\times\big\lbrace\delta(Q_3,Q_4)-\delta_0\delta(Q_3)\delta(Q_4)\big\rbrace \Big)\,.\nonumber
\end{eqnarray}
In the regimes SSBa and SSBd, where only either $\alpha_0$ or $\delta_0$ is nonzero, the mass term for the boson with zero order parameter is kept in the truncation for the effective potential.

The parametrization we use for the frequency- and momentum-dependence of the bosonic propagators and the Yukawa couplings can be found in Appendix B of Ref. \onlinecite{fourpoint}. The sole difference between the truncation used here for the symmetric regime SYM and in Ref. \onlinecite{fourpoint} is that nonzero quartic bosonic couplings $\bar\lambda_a$, $\bar\lambda_d$ and  $\bar\lambda_{ad}$ and a fermionic wave function renormalization factor $Z_F(\pi T)$ are taken into account in the present work. The presence of the quartic bosonic couplings is crucial for the flow in the symmetry broken regimes. In a purely fermionic language they correspond to vertices with eight fermions.

\section{Initial Conditions and Regulators}

As the microscopic scale $k=\Lambda$ goes to infinity, the flowing action must be equivalent to the microscopic action of the Hubbard model, so the initial value of the four-fermion coupling must correspond to the Hubbard interaction $U$. The bosonic fields decouple completely at this scale, where the initial values of the Yukawa couplings are given by
\begin{equation}
\bar h_a|_\Lambda=\bar h_\rho|_\Lambda=\bar h_s|_\Lambda=\bar h_d|_\Lambda=0\,.
\end{equation}
In practice, we choose a finite but very large $\Lambda$, which is a very good approximation.

For the bosonic mass terms we take $\bar m_{i,\Lambda}^2=t^2$ and $P_{i,\Lambda}=0$. The choice $\bar m_{i,\Lambda}^2=t^2$ amounts to an arbitrary choice for the normalization of the bosonic fields, which are introduced as redundant auxiliary fields at the scale $k=\Lambda$, where they do not couple to the electrons. Of course, this changes during the flow, where the bosons are transformed into dynamical composite degrees of freedom, with nonzero Yukawa couplings  and a nontrivial momentum dependence of their propagators. The quartic bosonic couplings vanish on initial scale $k=\Lambda$.

In addition to the truncation of the effective average action, regulator functions for both fermions and bosons have to be specified. We use ``optimized cutoffs'' \cite{litim1,litim2} for both fermions and bosons. The regulator function for fermions is given by
\begin{eqnarray}
R^F_k(Q)=\rm{sgn}(\xi(\mathbf q))\left( k-|\xi(\mathbf q)|\right)\Theta( k-|\xi(\mathbf q)|)\,,
\end{eqnarray}
while the regulator functions for the real bosons are given by
\begin{eqnarray}\label{regulator}
R^{a/\rho}_k(Q)=A_{a/\rho}\cdot(k^2/t^2-F_{c/i}(\mathbf q,\hat q))\Theta(k^2/t^2-F_{c/i}(\mathbf q,\hat q))
\,\end{eqnarray}
allowing for an incommensurability $\hat q$ with $F_{c/i}$ as defined in Appendix B of Ref. \onlinecite{fourpoint} (with $A_a=A_m$ and an additional $\Pi$-shift for the $a$-boson). Regulator functions for the Cooper-pair bosons are of the same form, but no incommensurability needs to be accounted for in these cases.

\section{Functional renormalization for the symmetric regime}

The flow equations for the couplings follow from projection of the exact flow equation for the effective average action onto the various different monomials of fields. The right hand sides of these flow equations are given by the 1PI diagrams having an appropriate number of external lines, including a scale derivative $\tilde \partial_k=(\partial_kR_k)\frac{\partial}{\partial R_k}$ acting only on the IR regulator $R_k$. Diagrams contributing to the flow of boson propagators and Yukawa couplings which do not include any quartic bosonic couplings have been discussed in Ref. \onlinecite{fourpoint}. Here we focus our discussion on diagrams which contribute to the flow of the quartic bosonic couplings $\bar\lambda_a$, $\bar\lambda_d$ and $\bar\lambda_{ad}$, and on how these couplings affect the flow of the bosonic mass terms and Yukawa couplings. We neglect the quartic couplings for the $s$- and $\rho$-bosons since the corresponding channels do not exhibit critical behavior in the parameter regimes studied.

\subsection{Bosonic mass terms}

\begin{figure}[t]
\includegraphics[width=55mm,angle=0.]{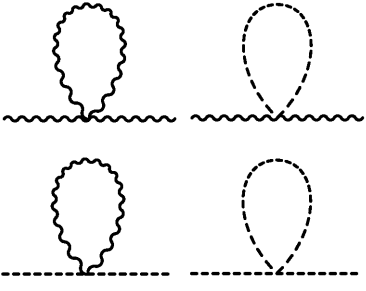}
\caption{\small{Diagrams involving the quartic bosonic couplings $\bar\lambda_a$, $\bar\lambda_d$ and $\bar\lambda_{ad}$ which contribute to the flow of the antiferromagnetic and $d$-wave superconducting propagators in SYM. Scale derivatives of the diagrams in the first line contribute to the flow of the propagator for the antiferromagnetic spin waves, those in the second line to the one for $d$-wave superconductivity. Wiggly lines denote antiferromagnetic, dashed lines superconducting bosons.}}
\label{propcorrections}
\end{figure}

The flow of the antiferromagnetic and $d$-wave superconducting mass terms is given by the following two equations,
\begin{eqnarray}\label{massflowa}
&&k\partial_k\bar m_a^2=2\bar h_a^2(0)\sum_Pk\tilde\partial_k\frac{1}{P_F^k(P)P_F^k(P+\Pi+\hat Q) }\\
&&-\sum_Pk\tilde\partial_k\left(\frac{5}{2}\frac{\bar\lambda_a}{P_a^k(P)+\bar m_a^2}+\frac{\bar\lambda_{ad}}{ P_d^k(P)+\bar m_d^2 }\right)\,\nonumber
\end{eqnarray}
and
\begin{eqnarray}\label{massflowd}
&&k\partial_k\bar m_d^2=-4\bar h_d^2(0)\sum_Pk\tilde\partial_k\frac{f_d(\mathbf p)^2}{P_F^k(P)P_F^k(-P) }\\
&&-\sum_Pk\tilde\partial_k\left(2\frac{\bar\lambda_d}{P_d^k(P)+\bar m_d^2}+\frac{3}{2}\frac{\bar\lambda_{ad}}{ P_a^k(P)+\bar m_a^2 }\right)\,.\nonumber
\end{eqnarray}
The first and second lines of these equations correspond to the fermionic and bosonic loop contributions, respectively. In Fig. \ref{propcorrections} we show a graphical representation of the contribution from bosons, while the fermionic diagrams can be found in Fig. 2 of Ref. \onlinecite{fourpoint}. The momentum vector $\hat Q$, which appears in the denominator of the fermionic contribution to $k\partial_k\bar m_a^2$ in Eq. \eqref{massflowa}, accounts for the dominance of \textit{incommensurate} over commensurate antiferromagnetic fluctuations in a wide range of parameters. In this case the minimum of the inverse antiferromagnetic propagator $\tilde P_a(Q)+\bar m_a^2$ no longer occurs for $Q=0$. Rather, there exist four discrete minima at vectors $\pm\hat Q_{x}=\pm(0,\hat q,0)$ and $\pm\hat Q_{y}=\pm(0,0,\hat q)$, either of which can be used as the vector $\hat Q$ in Eq. \eqref{massflowa}. For a detailed description of how incommensurate antiferromagnetism is treated within the present approach see Ref. \onlinecite{simon}.

\begin{figure}[t]
\includegraphics[width=70mm,angle=0.]{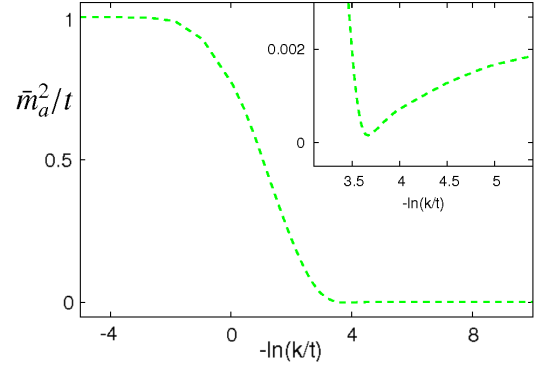}
\caption{\small{Flow of the antiferromagnetic mass term $\bar m_a^2$ for $U/t=3$, $t'/t=-0.1$, $\mu/t=-0.77$ and $T/t=0.0215$. The inset shows a detail of the flow where $\bar m_a^2$ reaches its minimal value, followed by an increase due to the bosonic contributions in the second line of Eq. \eqref{massflowa}.}}
\label{mainversion}
\end{figure}

In the symmetric regime, the fermionic contributions decrease the bosonic mass terms during the flow, whereas the bosonic contributions, proportional to the quartic couplings $\bar\lambda_a$, $\bar\lambda_d$ and $\bar\lambda_{ad}$, tend to increase them for positive $\bar\lambda_a$, $\bar\lambda_d$ and $\bar\lambda_{ad}$. The closer the mass terms approach zero, the more important the bosonic fluctuations become. Once the $\mathbf a$- or $d$- boson mass term becomes close to zero, bosonic fluctuations become relevant, and the terms in the second lines of Eqs. \eqref{massflowa} and \eqref{massflowd} may prevent it from actually reaching zero. Fig. \ref{mainversion} shows this by means of an example where the bosonic contribution proportional to $\bar\lambda_a$ inverts the direction of the flow of the mass term $\bar m_a^2$ so that it remains nonzero for $k\rightarrow0$.

Whenever a bosonic mass term $\bar m_i^2$ becomes zero during the flow, we change our description of the effective potential from the form of Eq. \eqref{Utrunc1} to that of Eq. \eqref{Utrunc2} or the corresponding versions for SSBa and SSBd. A negative quadratic term in the effective potential indicates local order, since at a given coarse graining scale $k$ the effective average action evaluated at constant field has a minimum for a nonzero value of the boson field. The largest temperature where at fixed values of $U,t',\mu$ one of the mass terms $\bar m_i^2$ vanishes during the flow is called the pseudocritical temperature $T_{pc}$. It can also be described as the largest temperature where short-range order sets in. At this temperature the effective momentum-dependent four-fermion coupling diverges in the channel where $\bar m_i^2$ hits zero, as seen in the purely fermionic flow studies like Ref. \onlinecite{halbothmetzner2}. However, the local order does not necessarily lead to long-range order, since the tendency toward order may be countered by long range bosonic fluctuations.

If the order persists for $k$ reaching a macroscopic scale, the model exhibits effectively spontaneous symmetry breaking, associated in our model to (either commensurate or incommensurate) antiferromagnetism or $d$-wave superconductivity. The true critical temperature $T_c$ is defined as the largest temperature for which local order persists up to some physical scale $k_{ph}$ corresponding to the inverse size of a macroscopic sample (see Refs. \onlinecite{bbw04,kw07}. We choose here $k_{ph}=(1cm)^{-1}\approx10^{-9}t$. In order to determine the true critical temperature for either $a$- or $d$-type of order, it is therefore necessary to switch to the truncation in which either $\alpha_0$ or $\delta_0$ (or both) are nonzero. Already a quick inspection of the phase diagram for$U/t=3$ and $t'/t=-0.1$ (Fig. \ref{phasediag}) reveals the importance of the flow in the spontaneously broken regimes. The pseudocritical temperature $T_{pc}$ differs substantially from the critical temperature $T_c$. In particular, the fact that local antiferromagnetic order is found at higher temperatures than superconducting order for $0.6<|\mu|/t<0.79$ does not imply that the system shows antiferromagnetic long-range order for these values of $\mu$. For low enough temperatures superconducting order actually prevails for $\mu/t>0.66$ The flow equations for the regimes with spontaneous symmetry breaking are discussed in more detail in the following section.

\begin{figure}[t]
\includegraphics[width=65mm,angle=0.]{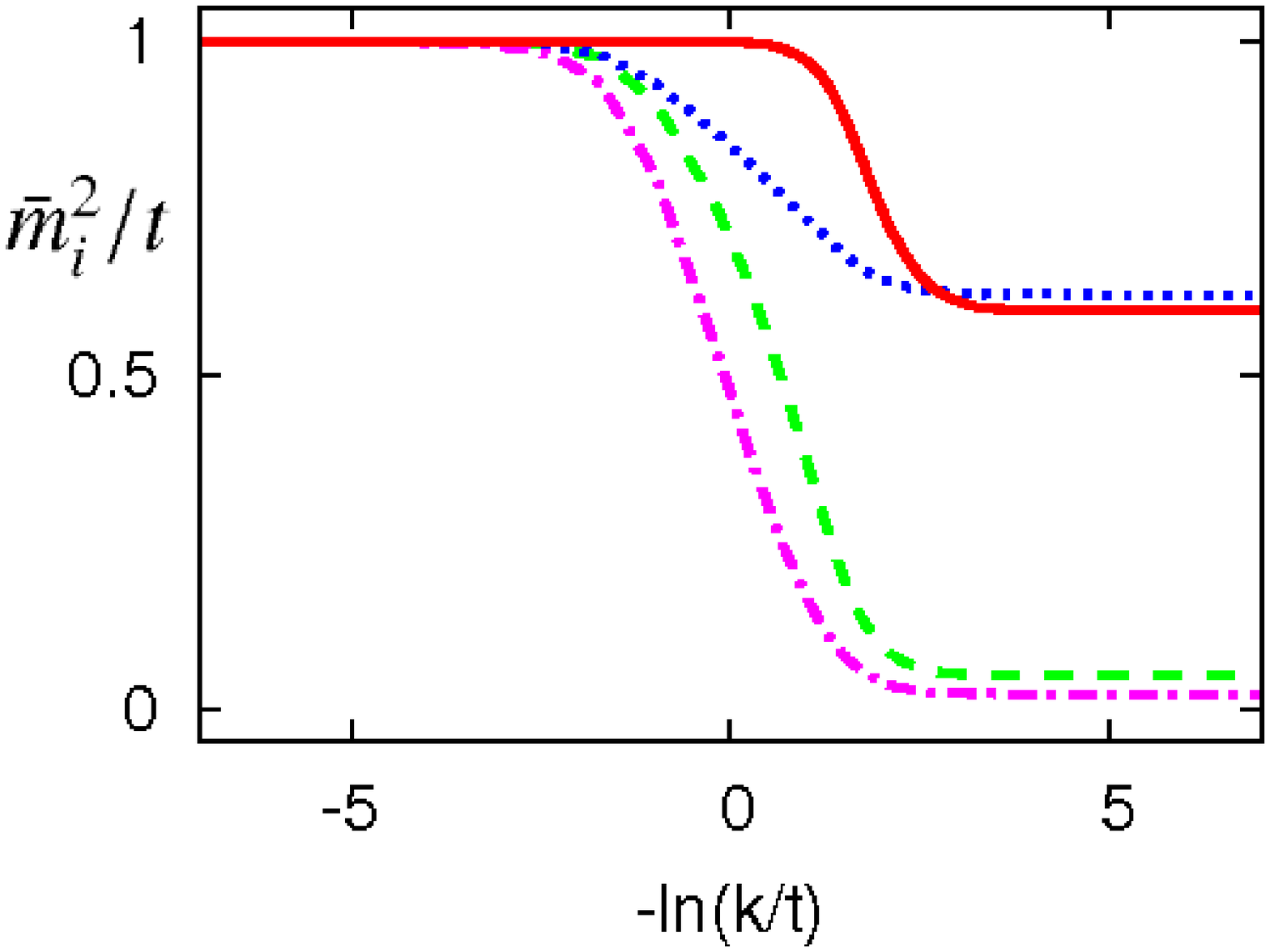}
\includegraphics[width=65mm,angle=0.]{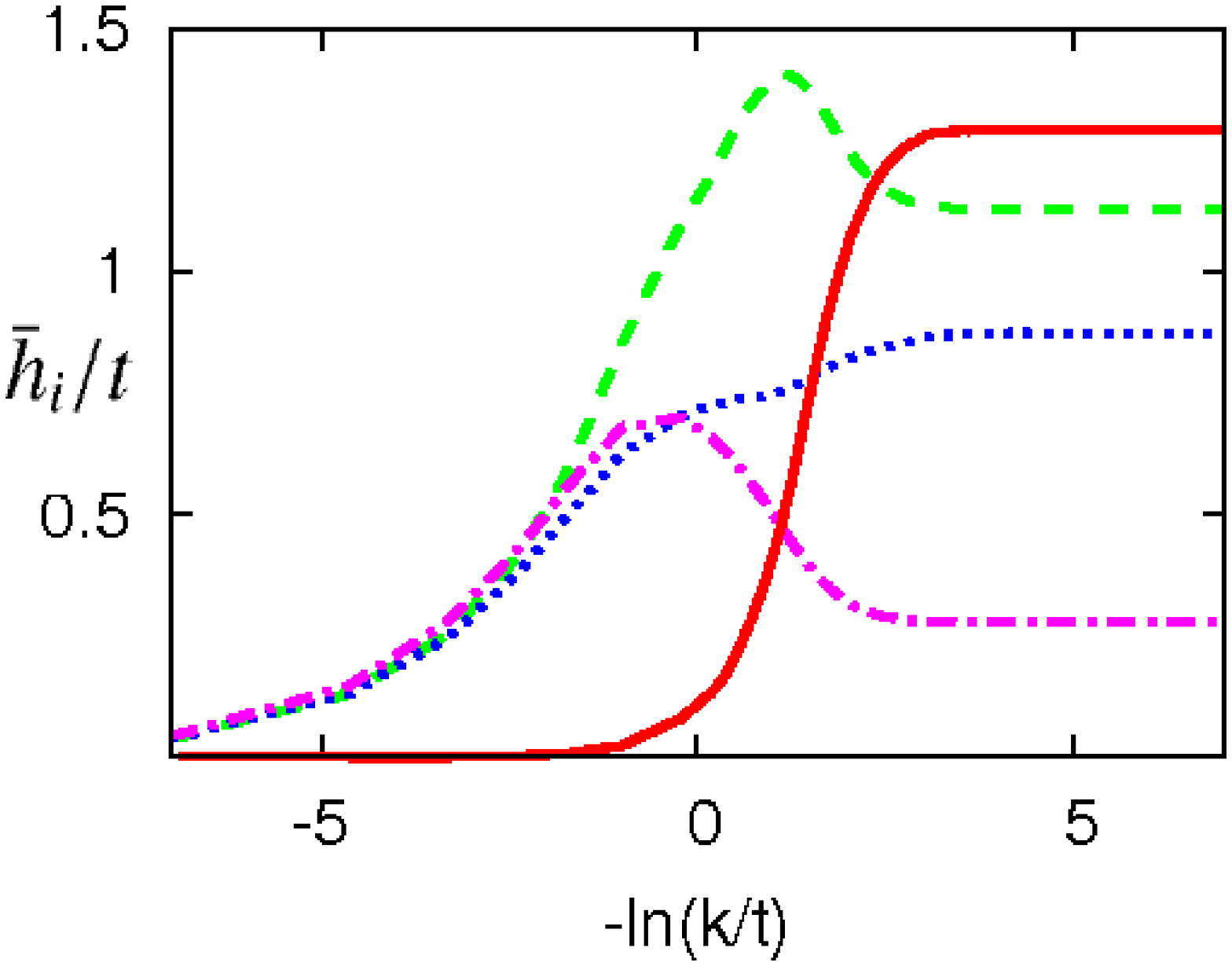}
\includegraphics[width=65mm,angle=0.]{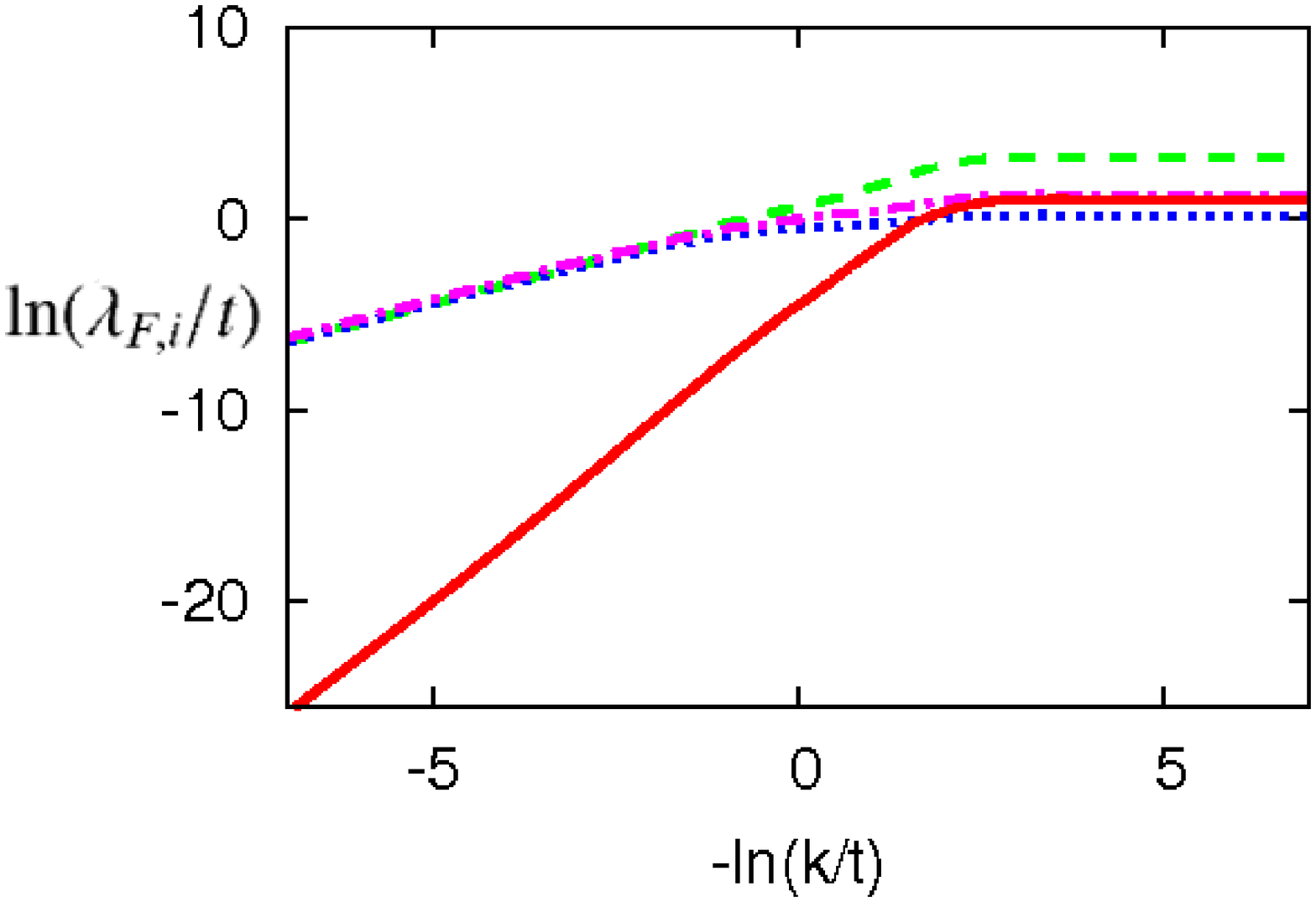}
\caption{\small{Flow of the bosonic mass terms $\bar m_a^2$, $\bar m_\rho^2$, $\bar m_s^2$  and $\bar m_d^2$ (upper panel), the Yukawa couplings $\bar h_a(0)$ , $\bar h_\rho(\Pi)$, $\bar h_s(0)$ and $\bar h_d(0)$ (middle panel). The lower panel shows a logarithmic plot of the effective fermionic four-point couplings $\lambda_{F,i}$ where $\lambda_{F,a}=\bar h_a^2(0)/\bar m_a^2$, $\lambda_{F,\rho}=\bar h_\rho^2(\Pi)/\bar m_\rho^2$, $\lambda_{F,s}=\bar h_s^2(0)/\bar m_s^2$ and $\lambda_{F,d}=\bar h_d^2(0)/\bar m_d^2$. The lines for all three panels are (green, dashed) for the antiferromagnetic boson, (red, solid) for the $d$-wave superconducting boson, (blue, dotted) for the charge density wave boson, and (magenta, dashed-dotted) for the $s$-wave superconducting boson. All dimensionful quantities are in units of $t$. Parameters chosen are $U/t=3$, $t'/t=-0.1$, $\mu/t=-0.6$ and $T/t=0.07$, where the system is always in the symmetric regime.}}
\label{massesyuks1}
\end{figure}

The flow of the bosonic mass terms and Yukawa couplings in the symmetric regime is shown in the first and second panels of Fig. \ref{massesyuks1}. Here we have chosen a chemical potential where antiferromagnetism is the dominant instability. Since the $s$-wave superconducting mass term falls slightly below the antiferromagnetic mass term and the Yukawa coupling in the $d$-wave channel $\bar h_d$ rises above the Yukawa coupling in the antiferromagnetic channel, one has to look at the ratios $\bar h_i^2/\bar m_i^2$ in order to see that the coupling in the antiferromagnetic channel is actually the dominant one. This is shown in the third panel of Fig. \ref{massesyuks1} where one can see that for the given choice of parameters the antiferromagnetic coupling is more strongly enhanced than the couplings in the $s$- and $d$-wave superconducting channels. The coupling in the charge density channel grows least of all four. We observe the very small value of the effective coupling for $d$-wave superconductivity at short distance scales (large $k$). This reflects the fact that this coupling is only generated by the antiferromagnetic fluctuations.

\begin{figure}[t]
\includegraphics[width=65mm,angle=0.]{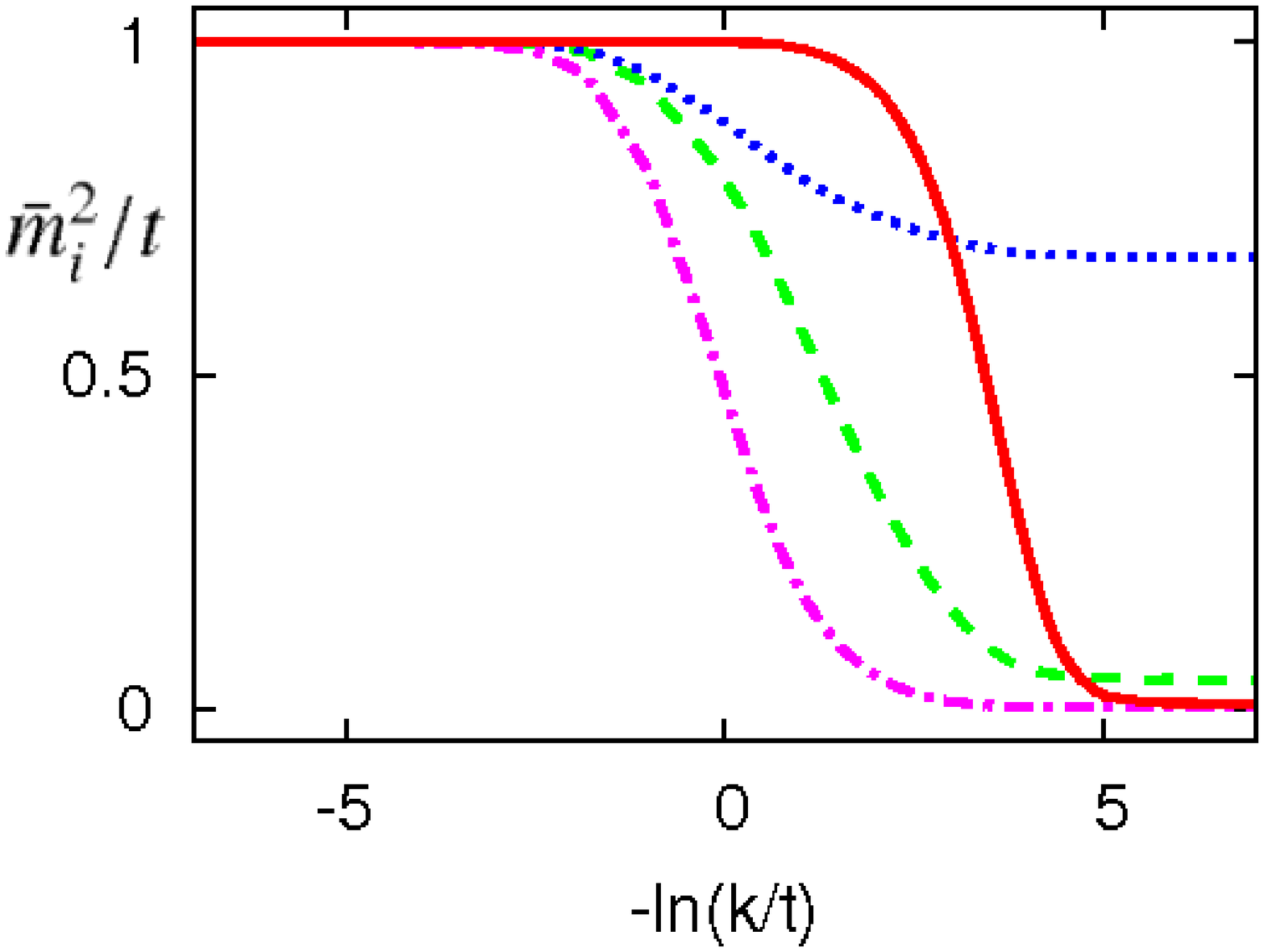}
\includegraphics[width=65mm,angle=0.]{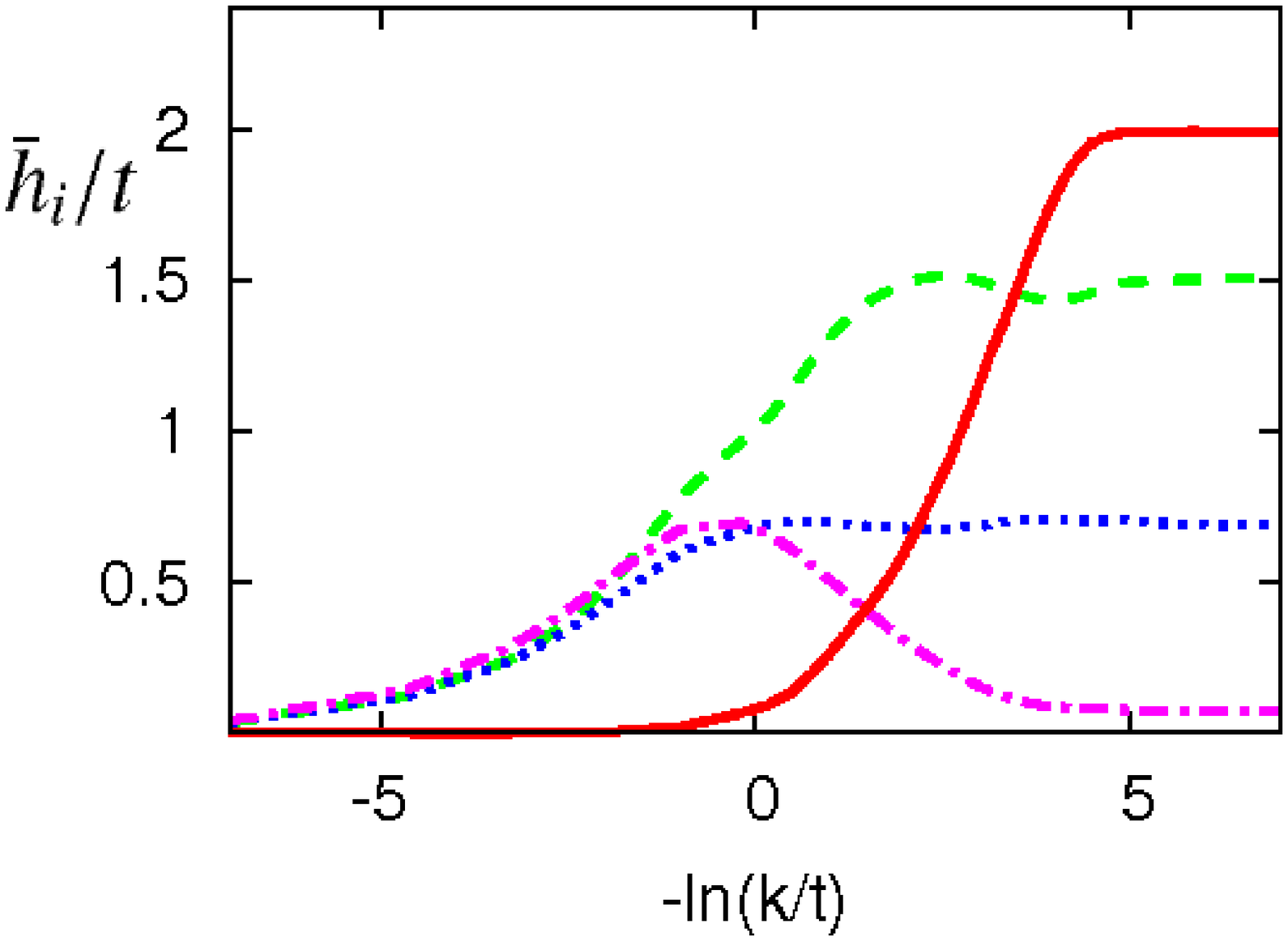}
\includegraphics[width=65mm,angle=0.]{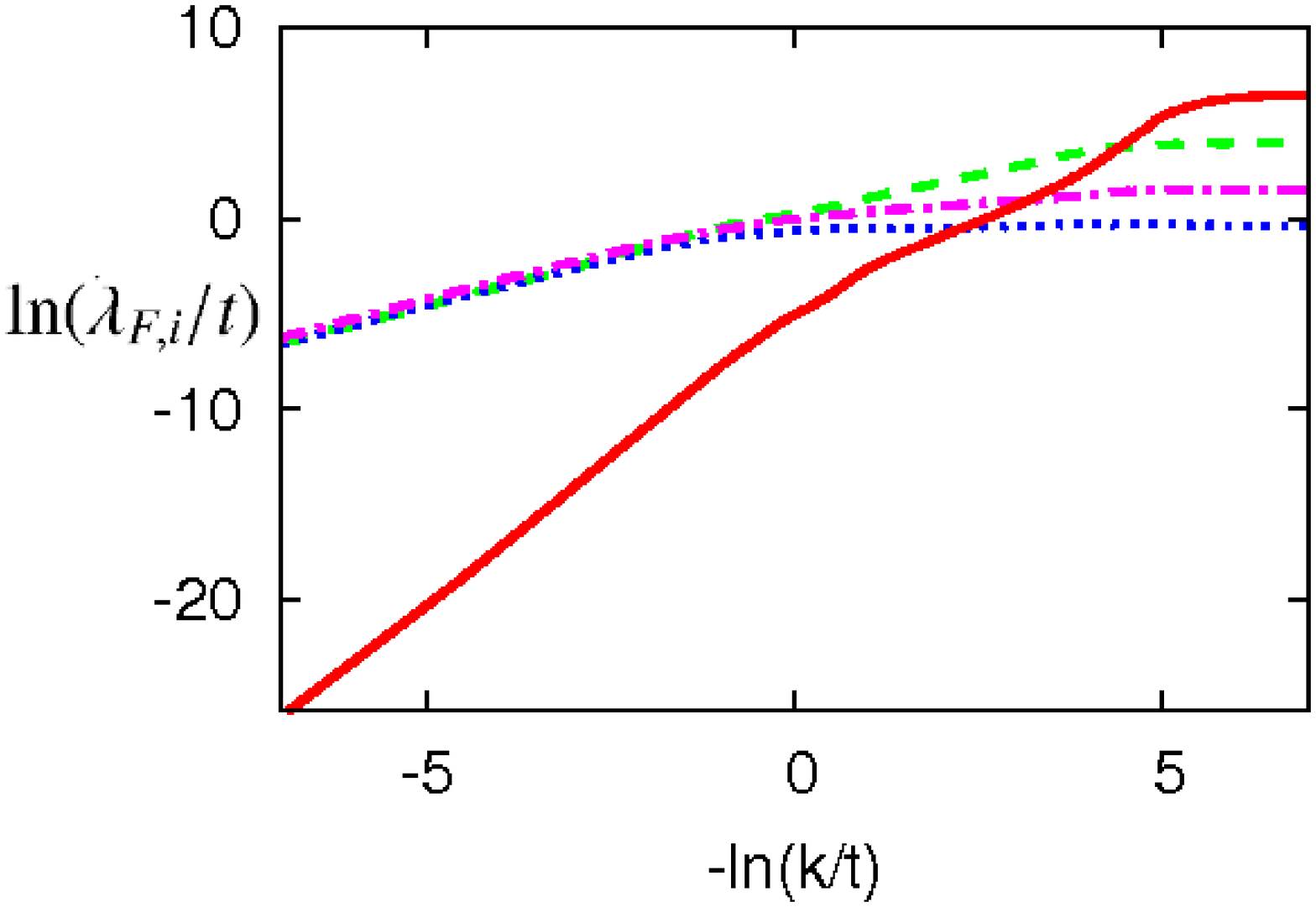}
\caption{\small{Same as Fig. \ref{massesyuks1}, now for $U/t=3$, $t'/t=-0.1$, $\mu/t=-0.83$ and $T/t=0.011$.}}
\label{massesyuks2}
\end{figure}

In Fig. \ref{massesyuks2} the flow of the bosonic mass terms, Yukawa couplings and effective fermionic four-point couplings is displayed for a combination of parameters where the coupling in the $d$-wave superconducting channel is the dominant one. Although this coupling is smallest on high scales of the flow by several orders of magnitude, it is strongly enhanced during the flow due to antiferromagnetic fluctuations, as discussed  within the context of the present framework in \cite{fourpoint}. At temperatures slightly lower than in Fig. \ref{massesyuks2} the mass term $\bar m_d^2$ reaches zero and the $d$-wave coupling diverges at a nonzero renormalization scale $k=k_{\rm{SSB}}$.

\subsection{Quartic bosonic couplings}

The flow of the quartic bosonic couplings $\bar\lambda_a$, $\bar\lambda_d$, and $\bar\lambda_{ad}$ is crucial for the long-range physics of the system in the symmetry-broken regimes, which is dominated by bosonic fluctuations. In order to obtain at $k=k_{SSB}$ the appropriate starting values for the flow of these couplings in the symmetry-broken regimes, however, one has to consider their flow already in the symmetric regime. If commensurate antiferromagnetic fluctuations dominate, the flow equation for the antiferromagnetic quartic coupling $\bar\lambda_a$ is given by
\begin{eqnarray}\label{lambdaaeq}
&&k\partial_k\bar\lambda_a=\Delta\dot{\Gamma}_{a}^{(4)}(0,0,0,0)\nonumber\\
&&=4\bar h_a^4(0)\sum_Pk\tilde\partial_k\frac{1}{\left( P_F^k(P)P_F^k(P+\Pi) \right)^2}\\
&&-\sum_Pk\tilde\partial_k\left(\frac{11}{2}\frac{\bar\lambda_a^2}{\left( P_a^k(P)+\bar m_a^2 \right)^2}+\frac{\bar\lambda_{ad}^2}{\left( P_d^k(P)+\bar m_d^2 \right)^2}\right)\,,\nonumber
\end{eqnarray}
where $\Delta\Gamma_{a}^{(4)}$ denotes the one-loop contribution to the bosonic four-point function, obtained as the fourth functional derivative of the flowing action with respect to the field $\mathbf a$, and the dot $^\cdot$ indicates the insertion of $k\tilde\partial_k$ under the measure of the loop integral implicit in $\Delta\Gamma_{a}^{(4)}$.  Where incommensurate fluctuations dominate over commensurate ones the flow equation \eqref{lambdaaeq} for $\bar\lambda_a$ has to be modified, yielding
\begin{eqnarray}
&&k\partial_k\bar\lambda_a=\frac{1}{2}\left(\Delta\dot{\Gamma}_{a}^{(4)}(\hat Q_x,-\hat Q_x,\hat Q_x,-\hat Q_x)\right.\nonumber\\
&&\hspace{2cm}\left.+\Delta\dot{\Gamma}_{a}^{(4)}(\hat Q_x,-\hat Q_x,\hat Q_y,-\hat Q_y)\right)\,.
\end{eqnarray}
For the quartic coupling $\bar\lambda_d$ of the $d$-boson one has the flow equation
\begin{eqnarray}
&&k\partial_k\bar\lambda_d=16\bar h_d^4(0)\sum_Pk\tilde\partial_k\frac{f_d(\mathbf p)^4}{\left( P_F^k(P)P_F^k(-P) \right)^2}\\
&&-\sum_Pk\tilde\partial_k\left(5\frac{\bar\lambda_d^2}{\left( P_d^k(P)+\bar m_d^2 \right)^2}+\frac{3}{2}\frac{\bar\lambda_{ad}^2}{\left( P_a^k(P)+\bar m_a^2 \right)^2}\right)\,.\nonumber
\end{eqnarray}

\begin{figure}[t]
\includegraphics[width=70mm,angle=0.]{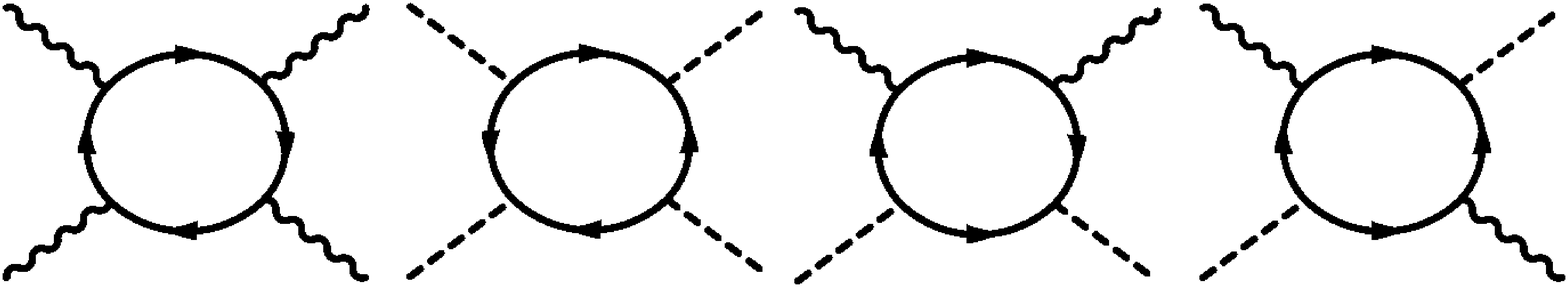}\vspace{0.5cm}
\includegraphics[width=70mm,angle=0.]{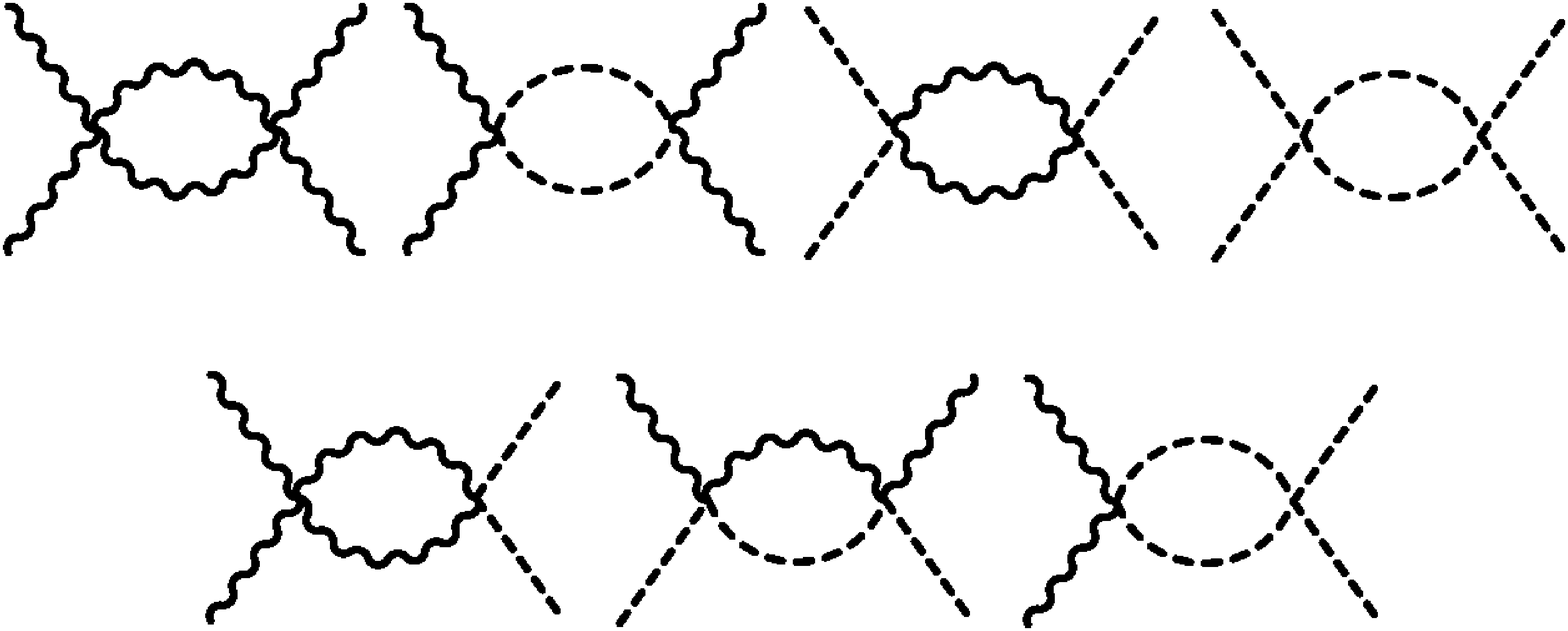}
\caption{\small{Contributions to the flow of the quartic bosonic couplings $\bar\lambda_a$, $\bar\lambda_d$ and $\bar\lambda_{ad}$ in SYM. The first line shows the contributions from fermionic, the second and third lines those from bosonic loops. Wiggly lines denote antiferromagnetic, dashed lines superconducting propagators, so the diagrams with four external wiggly lines contribute to the flow of $\bar\lambda_a$, those with four external dashed lines contribute to the flow of $\bar\lambda_d$ and those with two external wiggly and two external dashed lines to the flow of $\bar\lambda_{ad}$.}}
\label{quartcorrections}
\end{figure}

The flow equation for the quartic coupling $\bar\lambda_{ad}$ describing the mutual interaction between the $a$- and $d$-boson is given by
\begin{eqnarray}
&&k\partial_k\bar\lambda_{ad}=8\bar h_a^2(0)\bar h_d^2(0)\sum_Pk\tilde\partial_k\left(\frac{-2f_d(\mathbf p)^2}{\left( P_F^k(P)\right)^2P_F^k(-P)P_F^k(P+\Pi)}\right.\nonumber\\
&&\left.\hspace{1.5cm}+\frac{f_d(\mathbf p)f_d(\mathbf p+\mathbf\pi)}{ P_F^k(P)P_F^k(-P)P_F^k(P+\Pi)P_F^k(-P+\Pi)}\right)\nonumber\\
&&-\sum_Pk\tilde\partial_k\left(\frac{5}{2}\frac{\bar\lambda_a\bar\lambda_{ad}}{\left( P_a^k(P)+\bar m_a^2 \right)^2}+2\frac{\bar\lambda_d\bar\lambda_{ad}}{\left( P_d^k(P)+\bar m_d^2 \right)^2}\right. \nonumber\\
&&\hspace{1.5cm}\left.+2\frac{\bar\lambda_{ad}^2}{\left( P_a^k(P)+\bar m_a^2 \right)\left( P_d^k(P)+\bar m_d^2 \right)}\right)\,.
\end{eqnarray}
Graphical representations of the diagrams from which the contributions to the flow of $\bar\lambda_a$, $\bar\lambda_d$ and $\bar\lambda_{ad}$ are obtained as scale derivatives are given in Fig. \ref{quartcorrections}.

\begin{figure}[h]
\includegraphics[width=63mm,angle=0.]{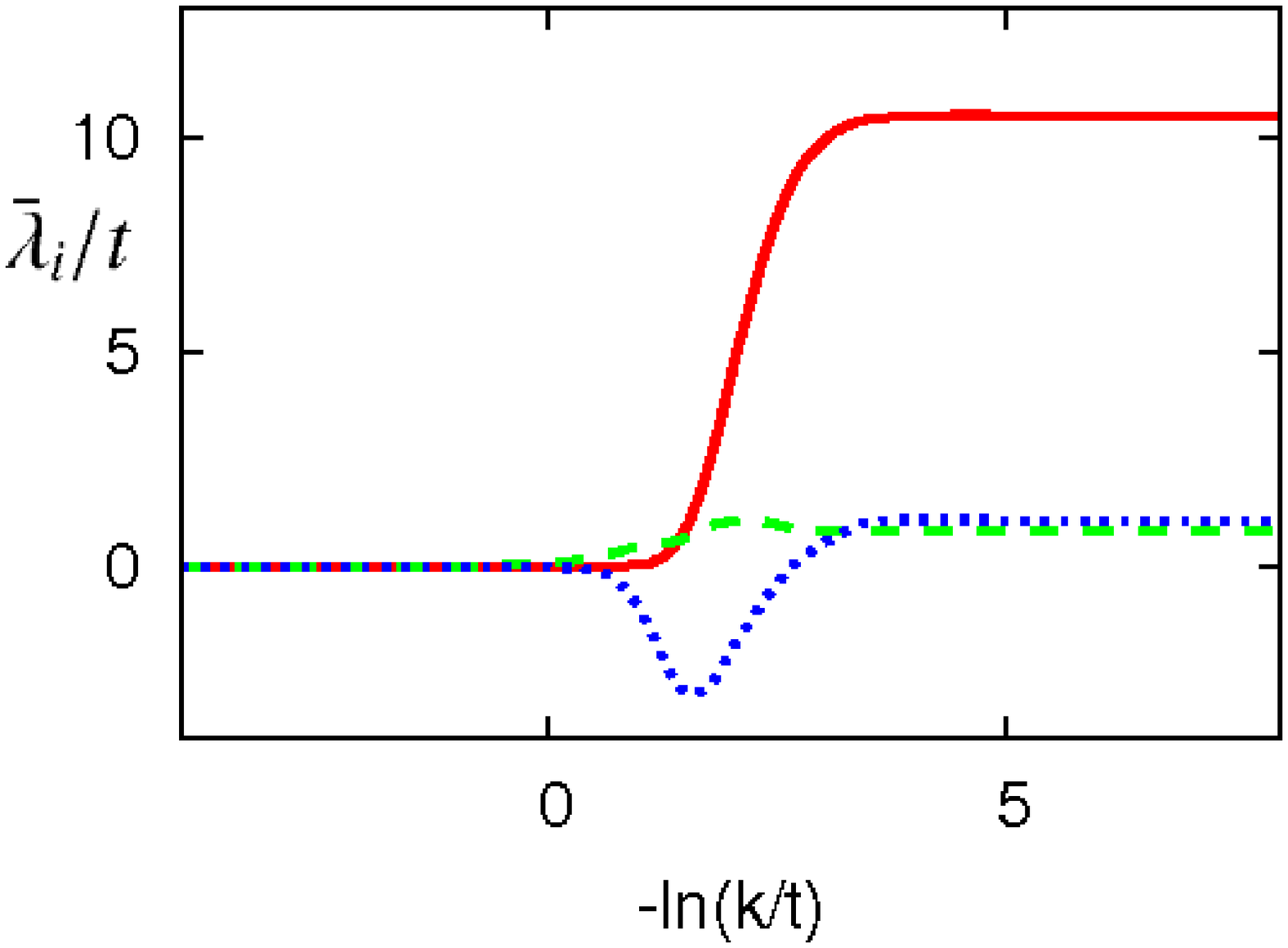}
\includegraphics[width=63mm,angle=0.]{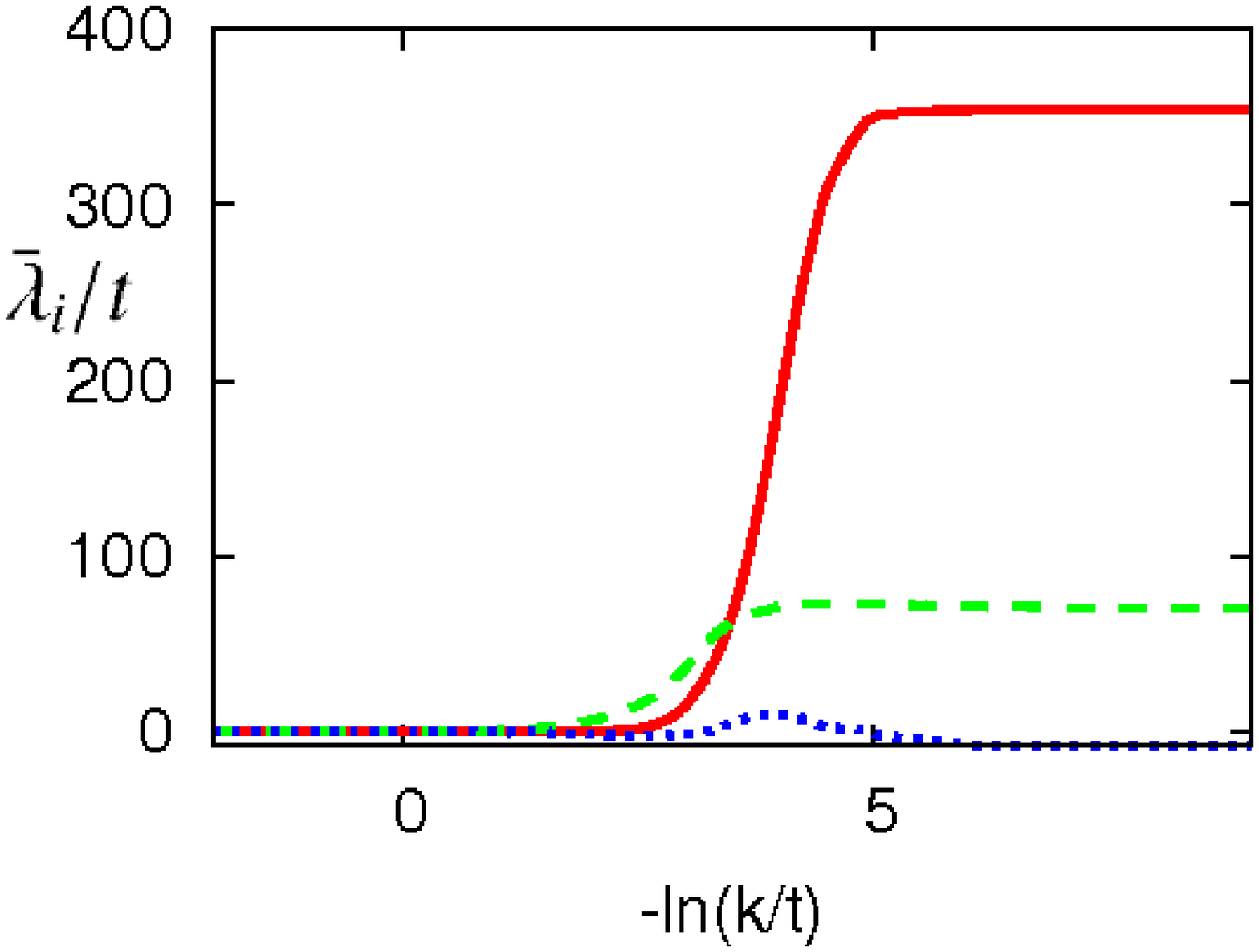}
\caption{\small{Upper panel: Flow of the (unrenormalized) quartic bosonic couplings $\bar\lambda_a$ (green, dashed), $\bar\lambda_d$ (red, solid) and $\bar\lambda_{ad}$ (blue, dotted) in SYM for $U/t=3$, $t'/t=-0.1$, $\mu/t=-0.6$ and $T/t=0.07$. Lower panel: The same for $\mu/t=-0.83$ and $T/t=0.011$, with $\bar\lambda_d$ multiplied by $0.1$.}}
\label{flowlambdas}
\end{figure}

The upper panel of Fig. \ref{flowlambdas} shows the flow of the quartic bosonic couplings $\bar\lambda_a$, $\bar\lambda_d$ and $\bar\lambda_{ad}$ for the same set of parameters as used in Fig. \ref{massesyuks1}. Although antiferromagnetism is the dominant instability for this choice of parameters, the quartic coupling $\bar\lambda_a$ (green, dashed curve) is only comparatively weakly enhanced during the flow. For smaller values of $-t'$ and not so close to half filling it may even turn negative during the flow so that the effective potential, according to the truncation \eqref{Utrunc1}, is no longer bounded from below so that the truncation is no longer adequate and has to be replaced by a more extended one. A negative value of $\bar\lambda_a$ may either indicate a tendency toward a first order antiferromagnetic phase transition, but it may also result from a more general inadequacy of the parametrization of the effective potential as a polynomial in the fields in the given range of parameters. To avoid these difficulties, which do not arise at larger values of $-\mu$ (see the green, dashed curve in the lower panel of Fig. \ref{flowlambdas}), we focus in this work on values of the parameters $t'$ and $\mu$ for which $\bar\lambda_a$ is non-negative on all scales.

While the coupling $\bar\lambda_a$ stays rather small during the flow and mostly has only a mild influence on the flow of the antiferromagnetic mass term in SYM, the quartic coupling $\bar\lambda_d$ can grow very large. Already for the parameters used in the upper panel of Fig. \ref{flowlambdas}, where the $d$-wave channel is far from critical, the coupling $\bar\lambda_d$ (red, solid curve) is substantially more enhanced than the quartic coupling $\bar\lambda_a$. The increase of $\bar\lambda_d$ is even stronger in the range of parameters where $d$-wave superconductivity is the dominant instability. This is shown in the lower panel of Fig. \ref{flowlambdas}, where $\bar\lambda_d$ is displayed after division by ten. The eminent growth of $\bar\lambda_d$ during the renormalization flow is chiefly responsible for the fact that the transition to $d$-wave superconductivity occurs only at rather large values of $-\mu$ as compared to the results in Ref.  \onlinecite{fourpoint} where no quartic bosonic couplings were taken into account.

The quartic coupling $\bar\lambda_{ad}$, which describes the direct interaction between the $\mathbf a$- and $d$- boson can change its sign from positive to negative, or inversely, during the renormalization flow, see Fig. \ref{flowlambdas} (short-dashed curves). If it is positive, it enhances the mass terms $\bar m_a^2$ and $\bar m_d^2$, otherwise it decreases them like the fermionic contributions to their flow.

\subsection{Anomalous dimensions and wave function renormalization}
For the long distance behavior of the system, the anomalous dimensions $\eta_a$ and $\eta_d$ are of importance. They are defined as
\begin{eqnarray}
\eta_a=-k\partial_k\ln A_a\;\hspace{1.2cm}\eta_d=-k\partial_k\ln A_d\,,
\end{eqnarray}
so they can be determined from the flow equations for $A_a$ and $A_d$. A description of how we access these quantities in the present approach can be found in Appendix B of Ref. \onlinecite{fourpoint}.

The flow equation for the fermionic wave function renormalization factor $Z_F=Z_F(\omega=\pm\pi T)$ is obtained from the flow of the fermionic propagator at the lowest two Matsubara modes $\pm\pi T$. We use the formula
\begin{eqnarray}\label{ZFflow}
k\partial_kZ_F=\frac{1}{2\pi iT}\,\left( \Delta\dot{\Gamma}^{(2)}_F(\pi T,\mathbf q_F)-\Delta\dot{\Gamma}^{(2)}_F(-\pi T,\mathbf q_F) \right)\,.
\end{eqnarray}
Here the subscript $F$ and the superscript $^{(2)}$ in $\Delta\Gamma^{(2)}_F$ indicate that the derivative has to be taken two times with respect to the fermionic fields. Again, the dot $^\cdot$ indicates the insertion of $k\tilde\partial_k$ under the measure of the loop integral implicit in $\Delta\Gamma_{F}^{(2)}$. In our ansatz some choice has to be made for the Fermi momentum $\mathbf q_F$ appearing on the right hand side of Eq. \eqref{ZFflow}. As we have checked, the increase of $Z_F$ during the flow is in general stronger for $\mathbf q_F$ close to the points $(0,\pm\pi)$ and $(\pm\pi,0)$ than for $\mathbf q_F$ close to the Brillouin zone diagonal \cite{honisalmiquasi}, but the precise choice does not matter for the semi-quantitative features of the phase diagram. For the results displayed in the figures we have set $\mathbf q_F=(0,\pi)$.

\begin{figure}[h]
\includegraphics[width=60mm,angle=0.]{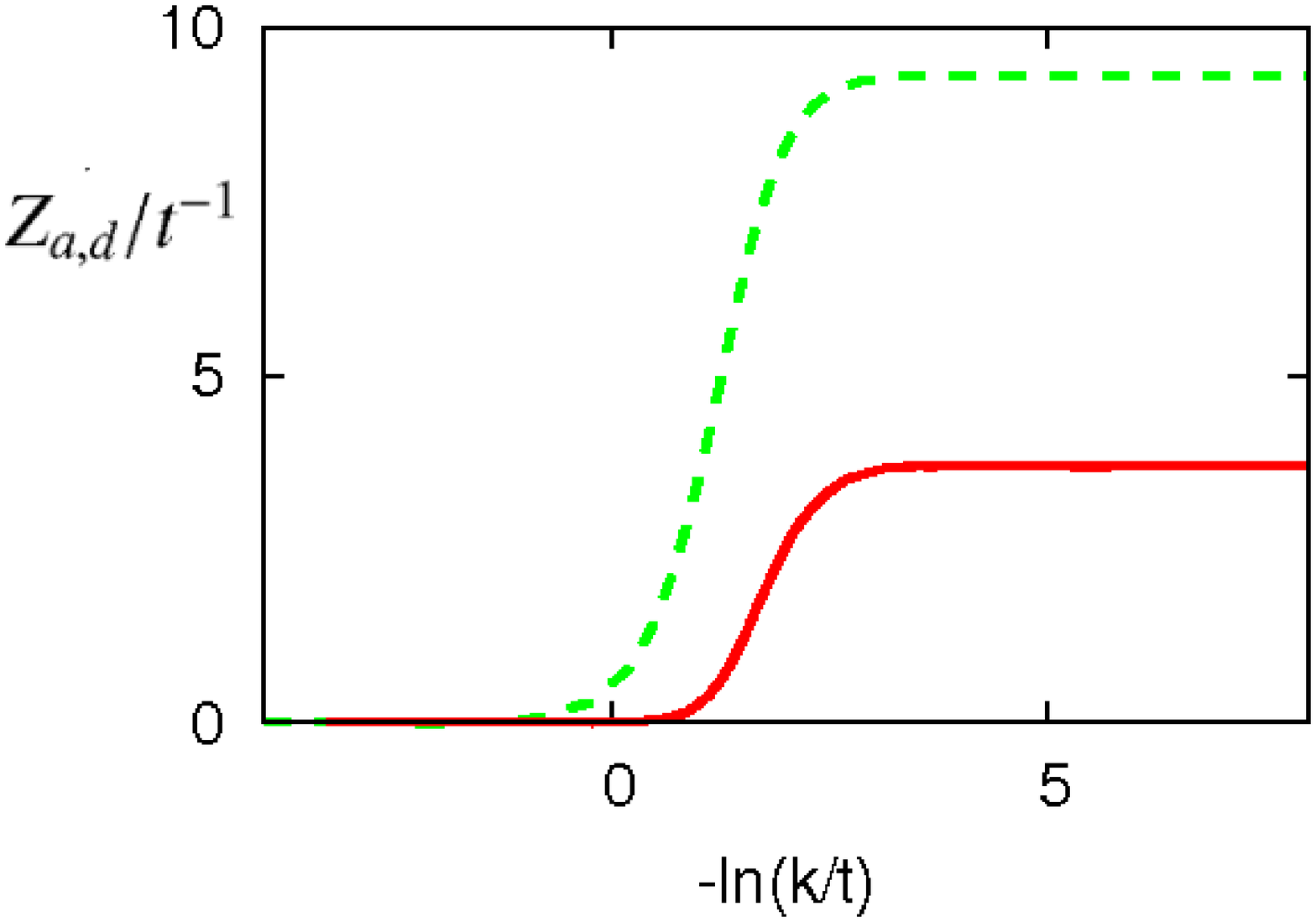}
\includegraphics[width=60mm,angle=0.]{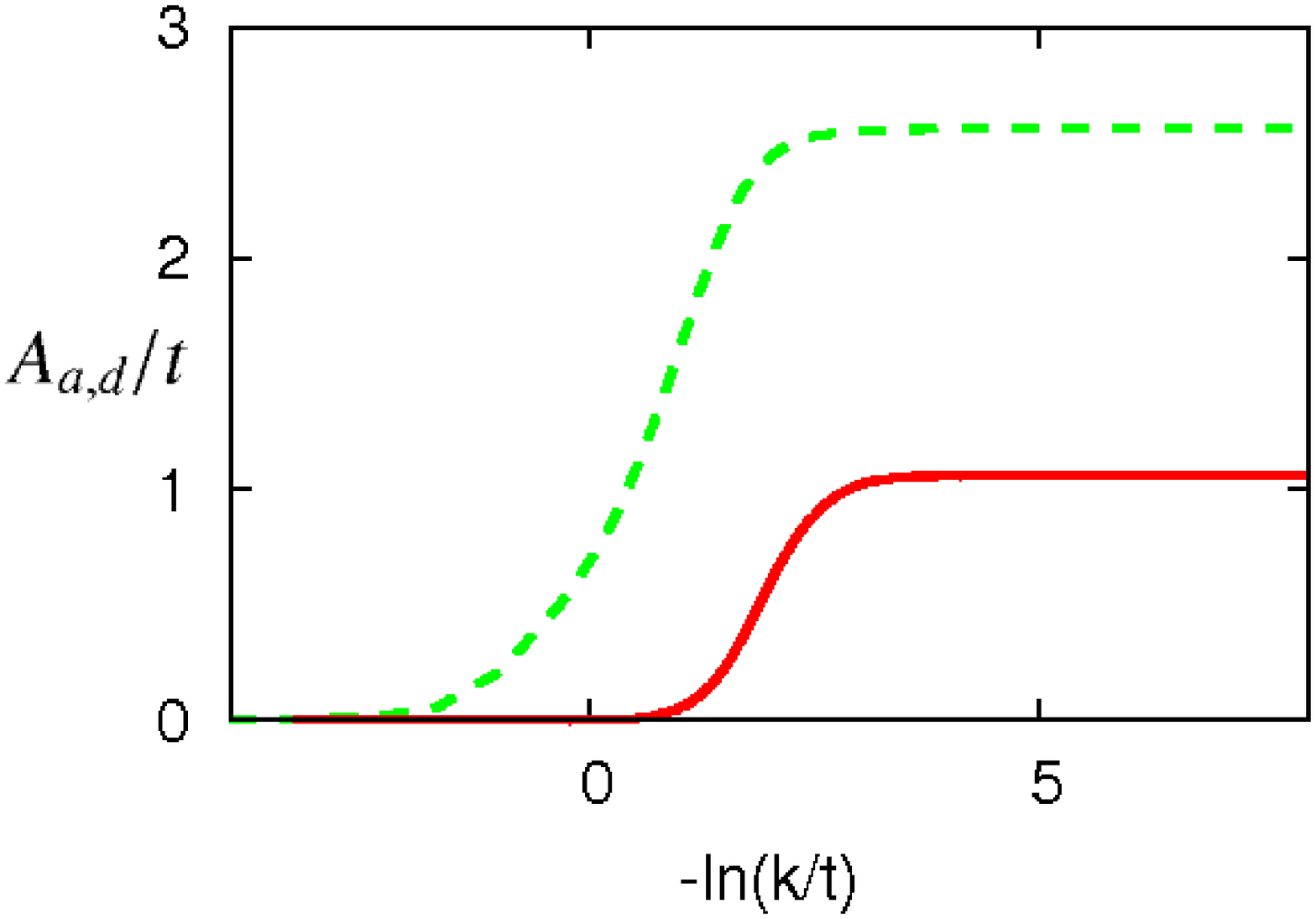}
\includegraphics[width=60mm,angle=0.]{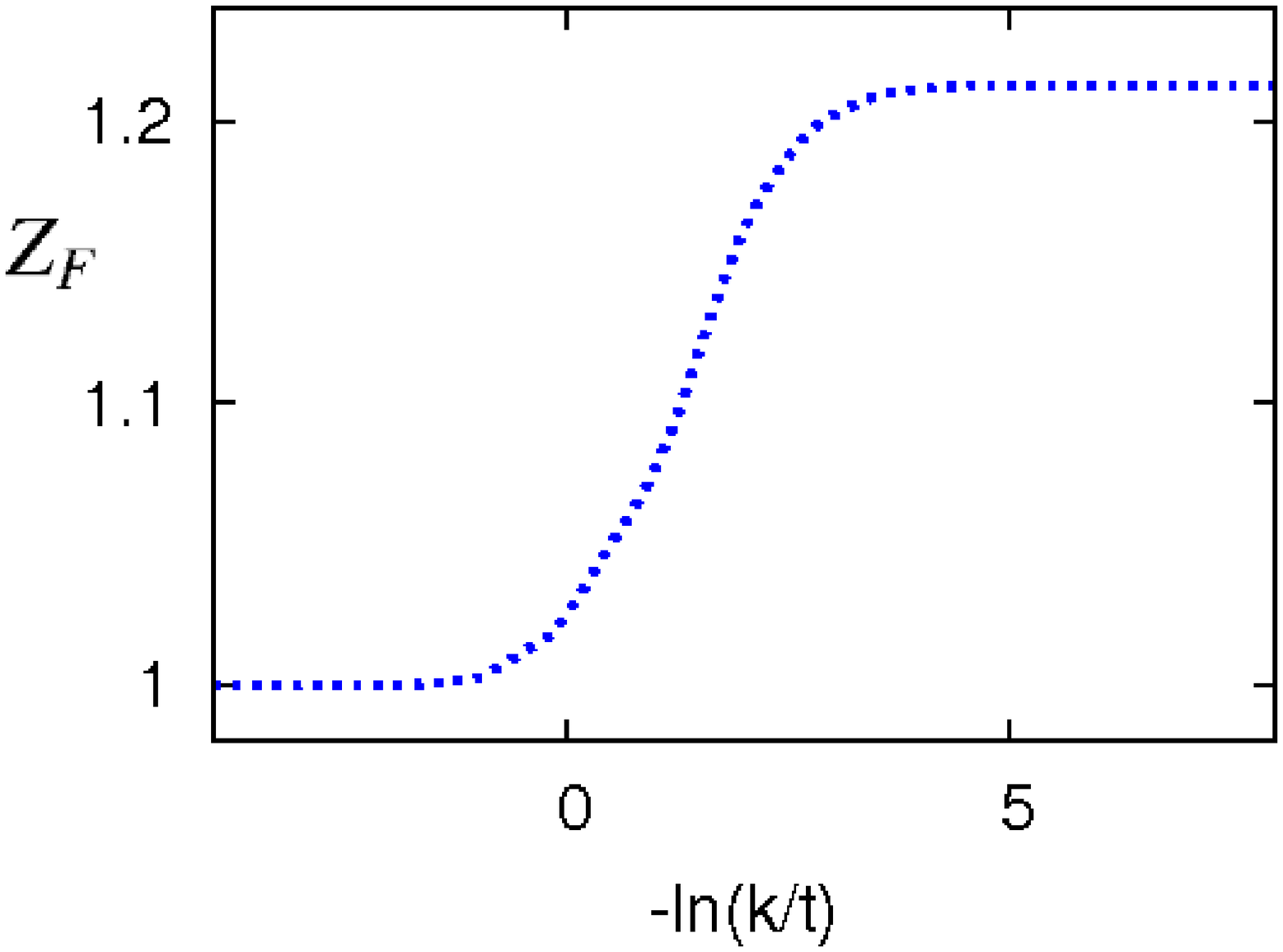}
\caption{\small{Upper panel: Flow of the bosonic wave function renormalization factors $Z_a$ (green, dashed) and $Z_d$ (red, solid). Middle panel: Flow of the gradient coefficients $A_a$ (green, dashed) and $A_d$ (red, solid). Lower panel: Flow of the fermionic wave function renormalization factor $Z_F(\pi T)$. All curves are for the symmetric regime at $U/t=3$, $t'=-0.1$, $\mu/t=-0.6$ and $T/t=0.07$.}}
\label{flowZs}
\end{figure}

The flow of the $Z$- and $A$-factors used in the parametrization of the $\mathbf a$- and $d$-boson propagators is displayed in the upper panels of Fig. \ref{flowZs}. The lower panel shows the fermionic wave function renormalization factor $Z_F(\pi T)$, which start its flow from $1$ and grows by some fraction for which the increase by $20\%$ in Fig. \ref{flowZs} is representative.

\section{Functional renormalization for the symmetry broken regimes}

In the studied parameter region we observe spontaneous symmetry breaking only in the antiferromagnetic and $d$-wave superconducting channels. For the range of $k$ where one of these channels shows local order we drop the charge density and $s$-wave superconducting bosons from our truncation and neglect the fermionic self-energy corrections. Furthermore, we restrict our attention to temperatures $T>T_{min}=4\cdot10^{-3}t$ and thus do not assess the ground state properties of the model at different values of $\mu$ and $t'$. Furthermore, we neglect the scale- and momentum-dependences of the Yukawa couplings $\bar h_a$ and $\bar h_d$, keeping their values fixed at those which they have at $k_{SSB}$: $\bar h_a|_k\equiv\bar h_a(0)|_{k_{SSB}}$ and $\bar h_d|_k\equiv\bar h_d(0)|_{k_{SSB}}$ for $k<k_{SSB}$. This neglect is made mainly due to computational reasons, but it should not have an important impact on the flow in the SSB-regimes at low temperatures, which is dominated by the long-range bosonic fluctuations, i.\ e. the bosonic masses, order parameters and quartic couplings.

As a final simplification, we neglect the incommensurability $\hat q$ in the SSB-regimes, which would otherwise have to be included in the truncation \eqref{Utrunc2} for the effective potential. Although including the incommensurability may have an effect on the flow of the antiferromagnetic order parameter at intermediate scales, we expect that it would not influence its flow at low scales, which is mainly determined by the number of Goldstone bosons. In addition to the continuous symmetry associated to the antiferromagnetic order parameter, incommensurate antiferromagnetic order breaks the symmetry of rotations of the lattice by $\pi/2$. Nevertheless, the spontaneous breakdown of this discrete symmetry does not lead to the emergence of additional Goldstone modes. We also do not think that the spontaneous breaking of lattice translation invariance results in major changes of the flow. Setting the incommensurability to zero in the spontaneously broken regimes will presumably leave the universal aspects of the flow of the running couplings in these regimes intact.

\subsection{Flowing potential for bosons}
For the SSB-regimes we derive the flow equations for the order parameters and quartic couplings from the flow equation of the local effective potential $U_B(\alpha,\delta)$ which is given by
\begin{equation}\label{effpoteq}
\partial_k U_B(\alpha,\delta)=\frac{1}{2}\rm{STr}\tilde\partial_k\ln\mathcal P_k[\alpha,\delta]\,.
\end{equation}
Here $\mathcal P_k[\alpha,\delta]$ corresponds to the cutoff-dependent full inverse propagator $\Gamma^{(2)}_k[\alpha,\delta]+R_k$ evaluated for constant bosonic fields. Within our truncation, the right hand side of the flow equation for the effective potential \eqref{effpoteq} can be decomposed into a fermionic and a bosonic contribution,
\begin{equation}
\partial_k U_B(\alpha,\delta)=\left(\partial_k U_B(\alpha,\delta)\right)^F+\left(\partial_k U_B(\alpha,\delta)\right)^B\,.
\end{equation}
The bosonic part can be written as
\begin{equation}\label{effboseq}
\left(\partial_k U_B[\alpha,\delta]\right)^B=\frac{1}{2}\sum_{P,i,j}\tilde\partial_k\ln\left[ P_i(P)\delta_{i,j}+\hat M_{i,j}^2(\alpha,\delta)+R_i^k(P)\delta_{i,j} \right]\,,
\end{equation}
where $P_i(P)=P_a(P)$ and $R_i(P)=R_a(P)$ for $i=1,2,3$, and $P_i(P)=P_d(P)$ and $R_i(P)=R_d(P)$ for $i=4,5$, respectively. The matrix $\hat M_{i,j}^2(\alpha,\delta)$, which has to be diagonalized, has entries
\begin{eqnarray}\label{Mmatrix}
\hat M_{i,j}^2(\alpha,\delta)=  \begin{cases}       \bar\lambda_a(3\alpha-\alpha_0)+\bar\lambda_{ad}(\delta-\delta_0),   &  i=j =1\,,\\
                                                    \bar\lambda_a(\alpha-\alpha_0)+\bar\lambda_{ad}(\delta-\delta_0),   &  i=j =2,3\,,\\
                                                    \bar\lambda_d(3\delta-\delta_0)+\bar\lambda_{ad}(\alpha-\alpha_0),   &  i=j =4\,,\\
                                                    \bar\lambda_d(\delta-\delta_0)+\bar\lambda_{ad}(\alpha-\alpha_0),   &  i=j =5\,,\\
                                                    \frac{1}{2}\bar\lambda_{ad}\sqrt{\alpha\delta},   &  i=1\; \&\; j=4\,,\\
                                                    \frac{1}{2}\bar\lambda_{ad}\sqrt{\alpha\delta},   &  i=4 \;\& \;j=1\,,\\
                                                    0 ,  &  \mbox{otherwise}\,.
             \end{cases}
\end{eqnarray}
The first and fourth lines and columns of the matrix $\hat M_{i,j}^2(\alpha,\delta)$ are associated to the radial, the others to the Goldstone modes. The radial modes of the two bosons are coupled to each other through the coupling $\bar\lambda_{ad}$ whereas the Goldstone modes remain unaffected. The form Eq. \eqref{Mmatrix} for the matrix $\hat M_{i,j}^2(\alpha,\delta)$ is adequate only in SSBad where the minimum of the effective potential $U(\alpha,\delta)$ occurs at nonzero values $\alpha_0$, $\delta_0$ of both order parameters $\alpha$ and $\delta$. For $\alpha_0=0$, the first three diagonal entries of $\hat M_{i,j}^2(\alpha,\delta)$ have to be replaced by $\bar m_a^2+3\bar\lambda_a\alpha$ (for $i=1)$ and $\bar m_a^2+\bar\lambda_a\alpha$ (for $i=2,3$). For $\delta_0=0$, the fourth and fifth diagonal entries of $\hat M_{i,j}^2(\alpha,\delta)$ have to be replaced by $\bar m_d^2+3\bar\lambda_d\delta$ and $\bar m_d^2+\bar\lambda_d\delta$.

The fermionic part $\left(\partial_k U_B(\alpha,\delta)\right)^F$ of the flow of the effective potential is given by
\begin{eqnarray}\label{effbfermeq}
\left(\partial_k U_B\right)^F=-\frac{1}{2}\rm{Tr}_F\tilde\partial_k\ln\mathcal P\,,
\end{eqnarray}
where the sum in the trace $\rm{Tr}_F$ is over fermionic indices only. Introducing the antiferromagnetic gap $\Delta_a^2=2\bar h_a^2\alpha_0$ and the (momentum-dependent) $d$-wave superconducting gap $\Delta_d^2(\mathbf q)=4\bar h_d^2f_d(\mathbf q)^2\delta_0$ this fermionic contributions to the flow of the effective potential can be derived from
\begin{eqnarray}
\left(\Delta U_B\right)^F&=&-\frac{1}{2}\rm{Tr}_F\ln\mathcal P\\
&=&-T\int_\mathbf p\frac{d^2p}{(2\pi)^2}\sum_{\epsilon=\{\pm1\}}\ln\,\cosh\left( \frac{\Theta_\epsilon}{2T} \right)\,,\nonumber
\end{eqnarray}
where
\begin{eqnarray}
\Theta_\epsilon &=&\Bigg\lbrack\left(\frac{1}{2} (\xi_{\mathbf p} + \xi_{\mathbf p+\boldsymbol\pi})+  \epsilon \sqrt{\frac{1}{4}(\xi_{\mathbf p} - \xi_{\mathbf p+\boldsymbol\pi})^2 + \Delta_a^2}\right)^2\nonumber\\
&&\hspace{2cm}+ \Delta_d^2(\mathbf q)\Bigg\rbrack^{1/2}\,.
\end{eqnarray}

By the help of Eqs.\ \eqref{effboseq} and \eqref{effbfermeq} the flow equations for the quartic couplings $\bar\lambda_a$, $\bar\lambda_d$ and $\bar\lambda_{ad}$ are obtained by appropriate derivatives with respect to the fields $\alpha$ and $\delta$ on both sides of Eq. \eqref{effpoteq},
\begin{eqnarray}\label{lambdaUeqs}
\partial_k\bar\lambda_a&=&\frac{d^2}{d\alpha^2}\left( \partial_kU(\alpha,\delta) \right)\big|_{\alpha=\alpha_0,\,\delta=\delta_0}\,,\nonumber\\
\partial_k\bar\lambda_d&=&\frac{d^2}{d\delta^2}\left( \partial_kU(\alpha,\delta) \right)\big|_{\alpha=\alpha_0,\,\delta=\delta_0}\,,\\
\partial_k\bar\lambda_{ad}&=&\frac{d^2}{d\alpha d\delta}\left( \partial_kU(\alpha,\delta) \right)\big|_{\alpha=\alpha_0,\,\delta=\delta_0}\,.\nonumber
\end{eqnarray}
These formulas are also valid if one of the symmetries remains unbroken in which case one has to set either $\alpha_0$ or $\delta_0$ to zero. For $\alpha_0=0$ or $\delta_0=0$ the flow of the mass term obeys
\begin{eqnarray}
\partial_k\bar m_a^2&=&\frac{d}{d\alpha}\left( \partial_kU(\alpha,\delta) \right)\big|_{\alpha_0=0,\,\delta=\delta_0}
\end{eqnarray}
or
\begin{eqnarray}
\partial_k\bar m_d^2&=&\frac{d}{d\delta}\left( \partial_kU(\alpha,\delta) \right)\big|_{\alpha=\alpha_0,\,\delta=0}\,.
\end{eqnarray}

\begin{figure}[t]
\includegraphics[width=70mm,angle=0.]{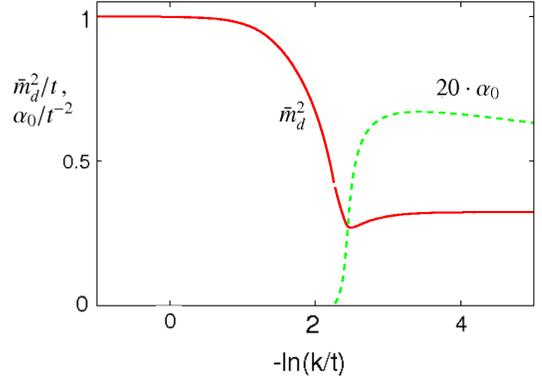}
\caption{\small{Flow of the $d$-wave superconducting mass term $\bar m_d^2$ (red, solid) and (unrenormalized) antiferromagnetic order parameter $\alpha_0$ (green, dashed), the latter multiplied by ten. Here nonzero $\alpha_0$ inverts the sign of the fermionic contribution to the flow of $\bar m_d^2$ so that it no longer decreases but rather increases and later saturates during the flow. Parameters chosen are $U/t=3$, $t'/t=-0.1$, $\mu/t=-0.65$ and $T/t=0.04$}}
\label{md_inversion}
\end{figure}

While in the symmetric regime the fermionic contributions to the flow of the mass terms are always negative and drive the masses toward zero, in the regimes with spontaneous symmetry breaking they may change sign. In this case even the fermionic contribution can lead to an \textit{increase} of the mass terms of the bosonic fields with vanishing order parameters. In particular, if the antiferromagnetic order parameter acquires a nonzero value, this may change the sign of the fermionic contribution to the flow of the superconducting mass term $\bar m_d^2$ and prevent it from becoming zero. This effect is shown in Fig. \ref{md_inversion}. In that sense the presence of antiferromagnetic order in the system has a tendency to prevent the establishment of $d$-wave superconducting order. Similarly, in the regimes where both $\alpha_0$ and $\delta_0$ are nonzero, a large value of $\alpha_0$ has a diminishing influence on the fermionic contribution to the flow of $\delta_0$, which therefore grows less quickly or decreases faster for $k\rightarrow0$ than if $\alpha_0$ were zero. This effect acts against the coexistence of antiferromagnetic and $d$-wave superconducting order. Indeed, the phase diagram in Fig. \ref{phasediag} shows no region of coexistence of both orders, in contrast to what one might have expected from the flow of the masses and Yukawa couplings in the symmetric regime and the pseudocritical temperatures.

\subsection{Flowing minimum}

To derive the flow equations of the order parameters $\alpha_0$ and $\delta_0$ we use the condition that $U_B(\alpha_0,\delta_0)$ should be the minimum of the effective potential $U_B(\alpha,\delta)$. From the necessary condition
\begin{eqnarray}
\partial_\alpha U_B(\alpha_0,\delta_0)=\partial_\delta U_B(\alpha_0,\delta_0)=0\,,
\end{eqnarray}
which has to hold on all scales, one obtains the prescription
\begin{eqnarray}
\frac{d}{dk}\partial_\alpha U_B(\alpha_0,\delta_0)=\frac{d}{dk}\partial_\delta U_B(\alpha_0,\delta_0)=0\,.
\end{eqnarray}
Together with Eq. \eqref{lambdaUeqs} the flow equations for the order parameters follow:
\begin{eqnarray}\label{orderUeqs}
\partial_k\alpha_0&=&-\frac{\bar\lambda_d}{\bar\lambda_a\bar\lambda_d-\bar\lambda_{ad}^2}\partial_\alpha\partial_kU_{B,k}(\alpha,\delta)\big|_{\alpha=\alpha_0,\,\delta=\delta_0}\nonumber\\ &&+\frac{\bar\lambda_{ad}}{\bar\lambda_a\bar\lambda_d-\bar\lambda_{ad}^2}\partial_\delta\partial_kU_{B,k}(\alpha,\delta)\big|_{\alpha=\alpha_0,\,\delta=\delta_0}\,,\nonumber\\
\partial_k\delta_0&=&-\frac{\bar\lambda_a}{\bar\lambda_a\bar\lambda_d-\bar\lambda_{ad}^2}\partial_\delta\partial_kU_{B,k}(\alpha,\delta)\big|_{\alpha=\alpha_0,\,\delta=\delta_0}\nonumber\\ &&+\frac{\bar\lambda_{ad}}{\bar\lambda_a\bar\lambda_d-\bar\lambda_{ad}^2}\partial_\alpha\partial_kU_{B,k}(\alpha,\delta)\big|_{\alpha=\alpha_0,\,\delta=\delta_0}\,.
\end{eqnarray}
For parameter regions where $\bar\lambda_a\bar\lambda_d-\bar\lambda_{ad}^2$ reaches zero the polynomial approximation for the flowing potential $U_B(\alpha,\delta)$ is no longer appropriate, and we discuss this issue below.

For the studied temperature regime $T>T_{min}$ the lowest Matsubara mode $n=0$ dominates in the spontaneously broken regimes ($k<k_{SSB}$) and the dimensionality of the problem is effectively reduced from $2+1$ to $2$, a mechanism which is known as ``dimensional reduction''. Within our flow equation approach dimensional reduction occurs automatically and in a smooth way due to the effective form of the flow equations \cite{tetradis}. It occurs effectively already for $k$ somewhat above $k_{SSB}$ where the contribution of bosons with $n\neq0$ becomes small. For computational simplicity, we therefore neglect contributions from all bosonic Matsubara modes except the lowest ones in the spontaneously broken regimes. This assumption becomes exact in the limit of $k\ll\pi T$, and we expect it to involve only a small quantitative inaccuracy at scales close to the critical scale. For very low temperature $T\lesssim k_{SSB}$ these arguments are no longer valid. This is the reason why the phase diagram in Fig. \ref{phasediag} is not shown for temperatures close to $T=0$. If $\alpha_0$ or $\delta_0$, but not both, are nonzero, the long-range behavior of the system at finite temperature can be described by the $O(3)$-symmetric \cite{halperinnelson} or $O(2)$-symmetric linear $\sigma$-model, depending on whether $\alpha_0$ or $\delta_0$ is nonzero. The properties of these models are well-known and well understood.

In a regime where the lowest Matsubara mode dominates over the others by far, one is dealing with an effectively two-dimensional problem, the unrenormalized field expectation values $\alpha_0$ and $\delta_0$ have to vanish in the infrared limit $k\rightarrow0$ in accordance with the Mermin-Wagner theorem. However, for the $O(2)$-symmetric model, which can be used as an approximation in the dimensionally reduced regime for superconducting order, the \textit{renormalized} field expectation value $\kappa_d=t^2A_d\delta_0/T$ may remain nonzero even if $\delta_0$ drops to zero as the gradient coefficient $A_d$ may diverge in this case \cite{cwls}. This behavior is characteristic of a Kosterlitz-Thouless phase transition \cite{kosterlitzthouless}, for functional renormalization group treatments see \cite{kw07,grater,gersdorff}. Although the polynomial expansion of the effective potential in Eq. \eqref{Utrunc2} is not sufficient to account for the finiteness of $\kappa_d$ down to $k=0$, it is sufficiently accurate to describe its being nonzero down to scales $k\ll k_{ph}$ much smaller than any realistic inverse probe size $l^{-1}$, see \cite{kw07} for a more detailed discussion.

In order to make contact with the familiar results from the $O(N)$-symmetric models, it is convenient to introduce the dimensionless (renormalized) quantities $\tilde\alpha$, $\kappa_a$, $\tilde\delta$, $\kappa_d$, $\lambda_a$, $\lambda_d$ and $\lambda_{ad}$, which are given by
\begin{eqnarray}
\tilde\alpha&=&\frac{t^2A_a}{T}\alpha,\;\hspace{1.2cm}\kappa_a=\frac{t^2A_a}{T}\alpha_0,\nonumber\\
\tilde\delta&=&\frac{t^2A_d}{T}\delta,\;\hspace{1.3cm}\kappa_d=\frac{t^2A_d}{T}\delta_0,\nonumber\\
\lambda_a&=&\frac{T}{t^2k^2A_a^2}\bar\lambda_a,\;\hspace{0.8cm}\lambda_d=\frac{T}{t^2k^2A_d^2}\bar\lambda_d,\nonumber\\
\lambda_{ad}&=&\frac{T}{t^2k^2A_aA_d}\bar\lambda_{ad}\,.
\end{eqnarray}
In terms of these quantities, when the fermions are fully gapped and dimensional reduction is efficient so that $k\ll2\pi T$ and $k\ll\pi$, the flow equations for the order parameters and quartic couplings at vanishing $\bar\lambda_{ad}=0$ reduce to those familiar from the $O(2)$- and $O(3)$-symmetric linear $\sigma$-models, namely
\begin{eqnarray}
k\partial_k\kappa_a&=&\frac{(4-\eta_a)}{16\pi}\left(\frac{3}{\left(1+2\lambda_a\kappa_a\right)^2}+2\right)-\eta_a\kappa_a\,,\label{kappa_a}\\
k\partial_k\lambda_a&=&\lambda_a^2\frac{(4-\eta_a)}{8\pi}\left(\frac{9}{\left(1+2\lambda_a\kappa_a\right)^3}+2\right)\nonumber\\
&&-2(1-\eta_a)\lambda_a\label{lambda_a}
\end{eqnarray}
for the $a$-boson, and
\begin{eqnarray}
k\partial_k\kappa_d&=&\frac{(4-\eta_d)}{16\pi}\left(\frac{3}{\left(1+2\lambda_d\kappa_d\right)^2}+1\right)-\eta_d\kappa_d\,,\label{kappa_d}\\
k\partial_k\lambda_d&=&\lambda_d^2\frac{(4-\eta_d)}{8\pi}\left(\frac{9}{\left(1+2\lambda_d\kappa_d\right)^3}+1\right)\nonumber\\
&&-2(1-\eta_d)\lambda_d\label{lambda_d}
\end{eqnarray}
for the $d$-boson. Since in the regime with two nonzero order parameters the absolute value of $\lambda_{ad}$ is normally driven to zero much faster than the two other quartic couplings $\lambda_a$ and $\lambda_d$, the flow of $\kappa_a$, $\kappa_d$, $\lambda_a$, $\lambda_d$ is generally well described by Eqs. \eqref{kappa_a}-\eqref{lambda_d}. If, however, $|\lambda_{ad}|$ is larger than the geometric mean of $\lambda_a$ and $\lambda_d$, that is if $|\lambda_{ad}|>\sqrt{\lambda_{a}\cdot\lambda_{d}}$, the effective potential $U(\alpha,\delta)$ no longer has a minimum at $(\alpha_0,\delta_0)$ and this signals the breakdown of our truncation which relies on an expansion of $U(\alpha,\delta)$ around $(\alpha_0,\delta_0)$, assumed to be the location of a minimum. Fortunately, however, our numerical results for the truncation Eq. \eqref{Utrunc2} yield a violation of the condition $|\lambda_{ad}|<\sqrt{\lambda_{a}\cdot\lambda_{d}}$ only in regions where antiferromagnetism strongly dominates over $d$-wave superconductivity. In this regime, the effect of $d$-wave superconducting fluctuations on the emergence of antiferromagnetic order is negligible and the truncation Eq. \eqref{Utrunc2} is not natural. Consequently, if in this regime $|\lambda_{ad}|$ rises above $\sqrt{\lambda_{a}\cdot\lambda_{d}}$, we set $\lambda_{ad}$ to zero on all scales whereby the expansion for the effective potential becomes again well-defined.

The main difference between the flow equations for $\kappa_a$ and $\lambda_a$ on the one hand and $\kappa_d$ and $\lambda_d$ on the other concerns the ``$+2$'' in Eqs. \eqref{kappa_a} and \eqref{lambda_a} as opposed to the ``$+1$'' in Eqs. \eqref{kappa_d} and \eqref{lambda_d}. This corresponds to the different numbers $2$ and $1$ of Goldstone bosons in the symmetry broken phases of the $O(3)$- and $O(2)$-symmetric linear $\sigma$-models, respectively. Since in the presence of a non-negligible order parameter the Goldstone modes have a much stronger influence than the radial modes in driving the order parameter to zero, their number is crucial for how long (in terms of the renormalization group flow) the system remains in the symmetry broken regime. The beta-functions for $\kappa_a$ and $\kappa_d$ are qualitatively different, since for $\kappa_d$ the contribution ``$+1$'' is canceled by the anomalous dimension, as we will see next.

\subsection{Anomalous dimensions}
In order to obtain the anomalous dimensions, one has to determine the flow equations for $A_a$ and $A_d$ in the presence of nonzero $\kappa_a$ and/or $\kappa_d$. To this end, we take a second derivative of the loop contributions to $P_a(Q)$ and $P_d(Q)$ with respect to spatial momentum and a derivative with respect to the scale $k$:
\begin{eqnarray}
\partial_kA_{a}=\partial_k\left(\lim_{l\rightarrow0}\frac{1}{2}\frac{\partial^2}{\partial l^2}\Delta P_{a}(0,l,0)\right)\,,\\
\partial_kA_{d}=\partial_k\left(\lim_{l\rightarrow0}\frac{1}{2}\frac{\partial^2}{\partial l^2}\Delta P_{d}(0,l,0)\right)\,.
\end{eqnarray}

In the regimes exhibiting spontaneous symmetry breaking, the fermionic contributions to $\eta_a$ and $\eta_d$ quickly become negligible as soon as the scale drops below the temperature, and it suffices to consider the bosonic contributions. In case the two bosons can independently be described by the two-dimensional $O(3)$- and $O(2)$-symmetric models these contributions are, assuming dimensional reduction,
\begin{eqnarray}\label{etasimple}
\eta_{a,d}=\frac{1}{\pi}\frac{\lambda_{a,d}^2\kappa_{a,d}}{\left(1+2\lambda_{a,d}\kappa_{a,d}\right)^2}.
\end{eqnarray}
In the presence of nonzero $\lambda_{ad}$, this formula has to be generalized, yielding
\begin{eqnarray}
\eta_a&=&\frac{1}{\pi}\left(\lambda_{ad}^2 \frac{\kappa_d (1 -  4 \kappa_a (\lambda_a - \kappa_d \lambda_{ad}^2 + 2 \kappa_d \lambda_a \lambda_d))}
{\left(1 + 2 \kappa_d \lambda_d +  2 \kappa_a (\lambda_a - 2 \kappa_d \lambda_{ad}^2 +  2 \kappa_d \lambda_a \lambda_d)\right)^2}\right.\nonumber\\
&&\left. +\frac{\kappa_a\lambda_a^2 (1 + 2 \kappa_d  \lambda_d)^2}
{\left(1 + 2 \kappa_d \lambda_d +  2 \kappa_a (\lambda_a - 2 \kappa_d \lambda_{ad}^2 +  2 \kappa_d \lambda_a \lambda_d)\right)^2}\right)\,,\\
\eta_d&=&\frac{1}{\pi}\left(\lambda_{ad}^2\frac{\kappa_a (1 - 4 \kappa_d(\lambda_d -\kappa_a \lambda_{ad}^2 + 2 \kappa_a \lambda_a \lambda_d)) }
{\left(1 +  2 \kappa_d \lambda_d +  2 \kappa_a (\lambda_a - 2 \kappa_d \lambda_{ad}^2 + 2 \kappa_d \lambda_a \lambda_d)\right)^2} \right. \nonumber\\
&&\left.+\frac{\kappa_d\lambda_d^2 (1 + 2 \kappa_a \lambda_a)^2}
{\left(1 +  2 \kappa_d \lambda_d +  2 \kappa_a (\lambda_a - 2 \kappa_d \lambda_{ad}^2 + 2 \kappa_d \lambda_a \lambda_d)\right)^2}\right)\,,
\end{eqnarray}
which reduces to Eq. \eqref{etasimple} for $\lambda_{ad}=0$.

\section{Temperature dependence of the order parameters}
\begin{figure}[t]
\hspace{-0.3cm}\includegraphics[width=64mm,angle=0.]{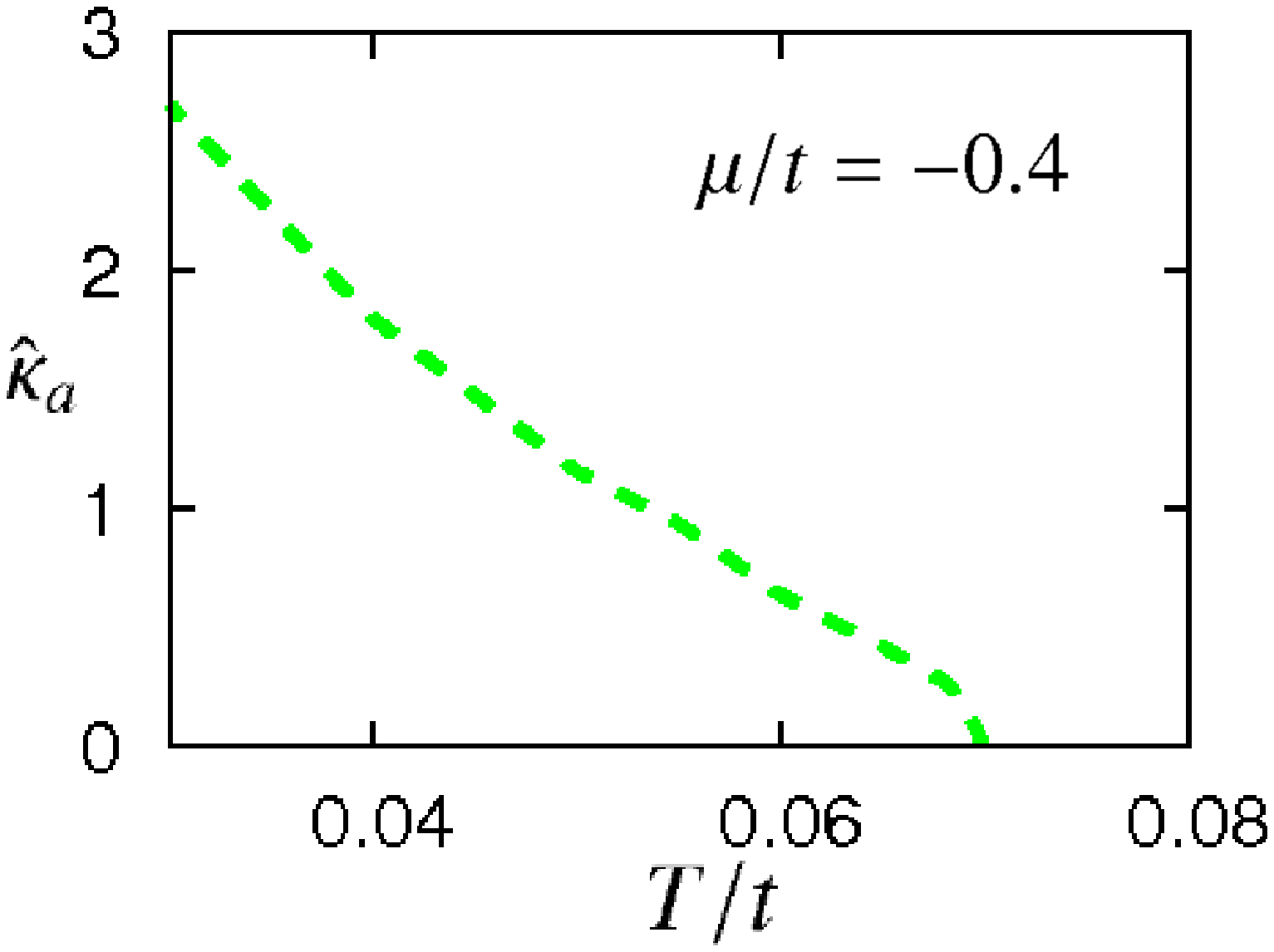}
\includegraphics[width=64mm,angle=0.]{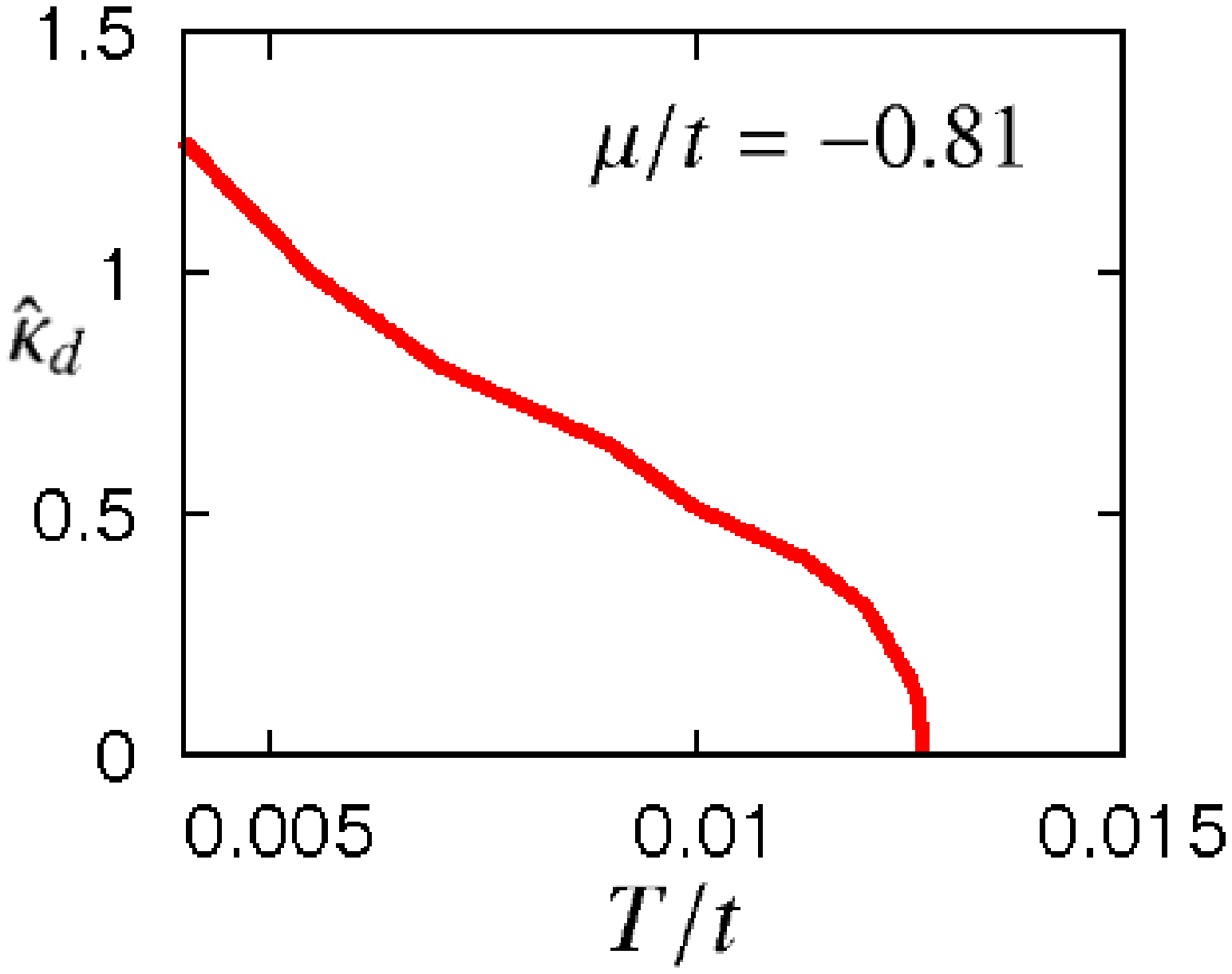}
\includegraphics[width=64mm,angle=0.]{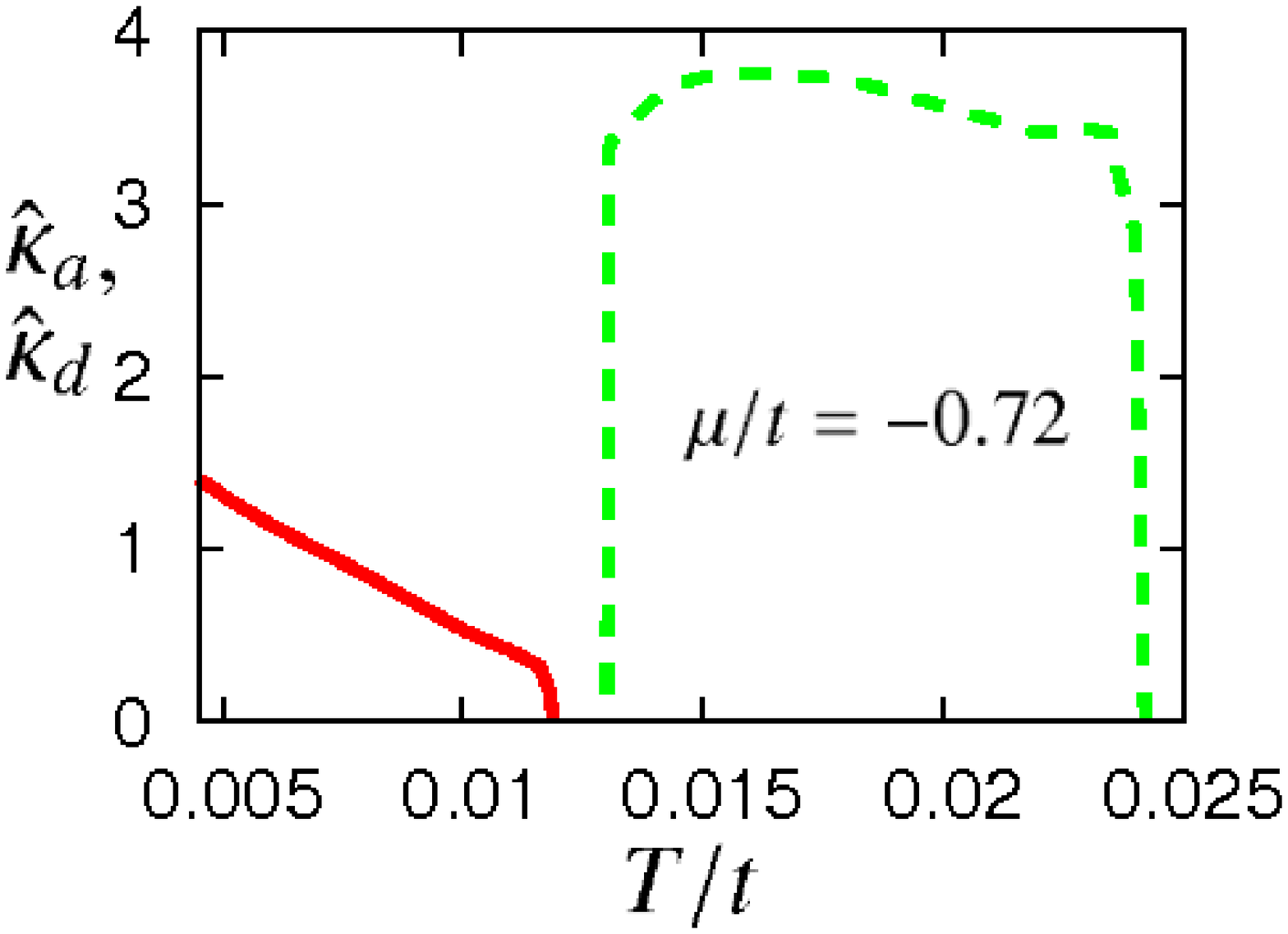}
\caption{\small{Renormalized order parameters $\hat\kappa_a$ and $\hat\kappa_d$ at the ``macroscopic'' scale $k=k_{ph}$, corresponding to an inverse probe size of $\approx1\ cm$, as a function of temperature $T$ for $U/t=3$ and different values of $\mu$. The upper panel shows the temperature dependence $\hat\kappa_a$ for $\mu/t=-0.4$, the middle panel shows $\hat\kappa_d$ for $\mu/t=-0.81$. The lower panel shows  $\hat\kappa_d$ and $\hat\kappa_a$ for $\mu/t=-0.72$, where they are nonzero in different temperature ranges.}}
\label{orderparams}
\end{figure}

We now turn to the discussion of our numerical results for the order parameters and the gaps as functions of temperature (Fig. \ref{orderparams}), the flow of the running couplings in a selected number of cases (Figs. \ref{order_parameters_1} and \ref{order_parameters_2}), and the more general features of the phase diagram (Fig. \ref{phasediag}) in the symmetry broken regimes SSBa, SSBd and SSBad.

\subsection{Order parameters}
Fig. \ref{orderparams} displays the renormalized antiferromagnetic and $d$-wave superconducting order parameters $\kappa_a$ and $\kappa_d$ as functions of temperature at different values of the chemical potential $\mu$ for $U=3t$ and $t'=-0.1t$. Both $\kappa_a$ and $\kappa_d$ are evaluated at $k_{ph}=10^{-9}t\approx1\,\rm{cm}^{-1}$, corresponding to a realistic inverse probe size. The upper panel shows the temperature dependence of $\hat\kappa_a=\kappa_a\big|_{k=k_{ph}}$ at the van Hove filling $\mu=4t'$. The shape of this curve for $\hat\kappa_a$ is similar to that of the curve presented Fig. 1 in Ref. \onlinecite{bbw04} for $t'=\mu=0$. The temperatures where $\hat\kappa_a$ is nonzero, however, are lower according to the results presented here since more fluctuations have been included which have a tendency to destroy antiferromagnetic order. The middle panel of Fig. \ref{orderparams} shows the temperature dependence of $\hat\kappa_d=\kappa_d\big|_{k=k_{ph}}$ at $\mu/t=-0.81$, where only $d$-wave superconducting order and no antiferromagnetic order occurs. The steep fall to zero of $\hat\kappa_d$ at $T=T_c$ can be seen as a remnant of the jump in the superfluid density found for a Kosterlitz-Thouless phase transition at $T_c$ in an improved truncation \cite{gersdorff,kw07}.

In the absence of electromagnetic interactions (as for the pure Hubbard model) the superconducting phase is actually a superfluid phase with superfluid density given by $n_s=\hat\kappa_d/a^2$, where $a$ denotes the lattice spacing and $n_s$ the number of particles per area. For nonzero electromagnetic coupling $e$ the ``photon mass'' responsible for superconductivity is given by $m_\gamma=2e\sqrt{\hat\kappa_d}$. (Here we observe that the $d$-boson carries charge two. More precisely, $e$ is the effective renormalized electromagnetic coupling at the scale $k_{ph}$.) Other observable quantities are the effective gaps for the electrons. Indeed, the antiferromagnetic and $d$-wave superconducting gaps $\Delta_a$ and $\Delta_d$ are related to $\hat\kappa_a$ and $\hat\kappa_d$ by $\Delta_a=\sqrt{2h_a^2\hat\kappa_a}$ and $\Delta_d(\mathbf q)=\sqrt{ h_d^2f_d(\mathbf q)^2\hat\kappa_d}$, where the renormalized Yukawa couplings $h_a$ and $h_d$ are given by $h_{a,d}^2=\frac{T}{A_{a,d}t^4}\bar h_{a,d}^2$. Fig. \ref{orderparams} therefore predicts measurable quantities.

The lower panel of Fig. \ref{orderparams} shows a situation where, at $\mu/t=-0.72$, we find nonzero $\kappa_d$ at low temperatures and nonzero $\kappa_a$ at higher temperatures. In between, we observe a small temperature region around $T=0.0125t$ where neither of the two order parameters remains nonzero down to $k=k_{ph}$. In this region only local but no long-range order is present. An interesting feature of this graph is the steepness of the rise of $\hat\kappa_a$ at $T=T_c$, which contrasts with the behavior at van Hove filling (upper panel of Fig. \ref{orderparams}) where the rise below $T_c$ is relatively smooth. The main reason for this feature is the strong growth with temperature of the final value of $\lambda_a$ in SYM close to $T=T_c$. This value has an important influence on the initial growth of $\kappa_a$ in the spontaneously broken regime and therefore at its value at $k=k_{ph}$. This effect is mainly responsible for the smallness of the temperature interval in which $\hat\kappa_a$ drops to zero as the temperature approaches $T_c$ from below.

\begin{figure}[t]
\includegraphics[width=64mm,angle=0.]{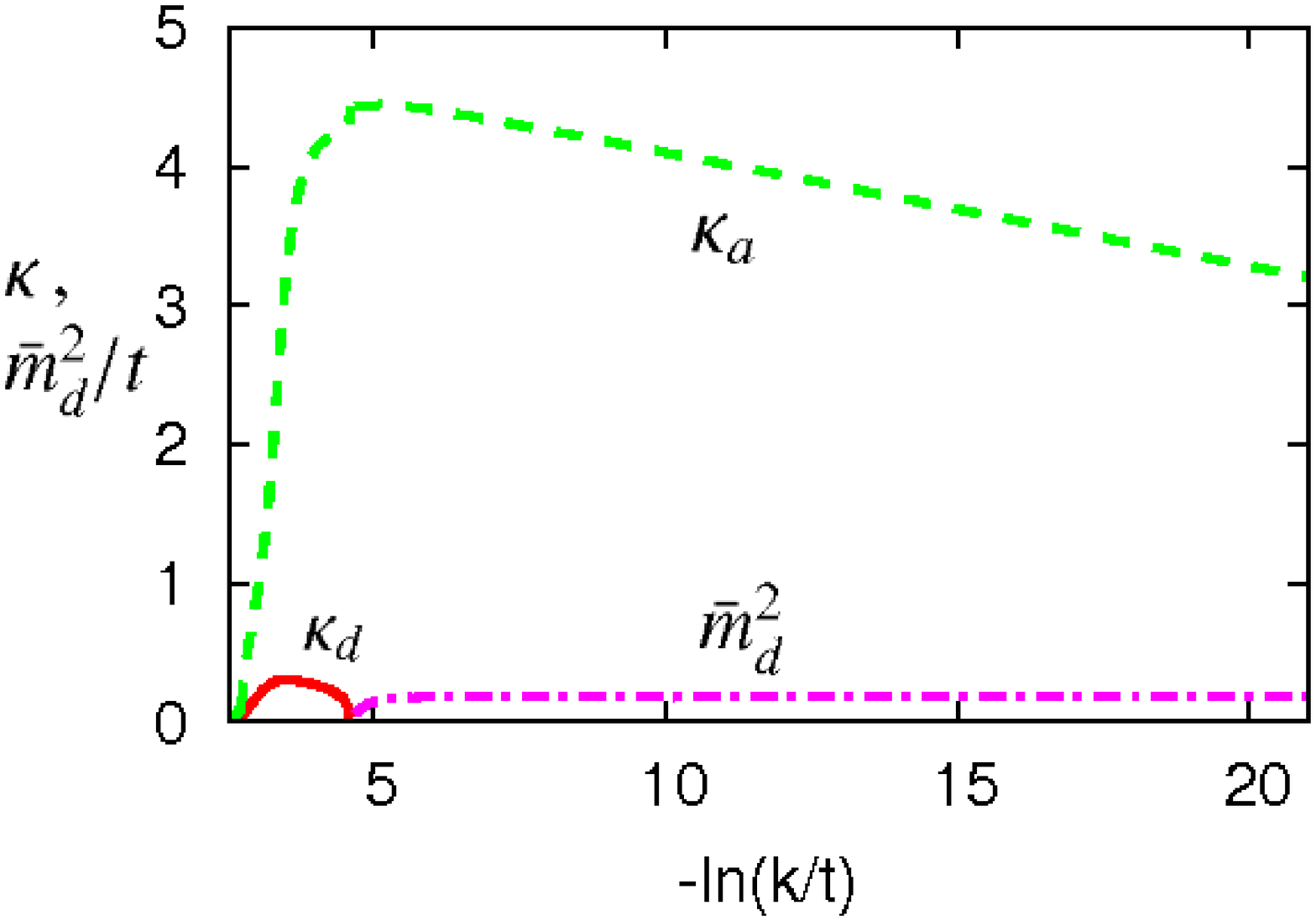}
\includegraphics[width=64mm,angle=0.]{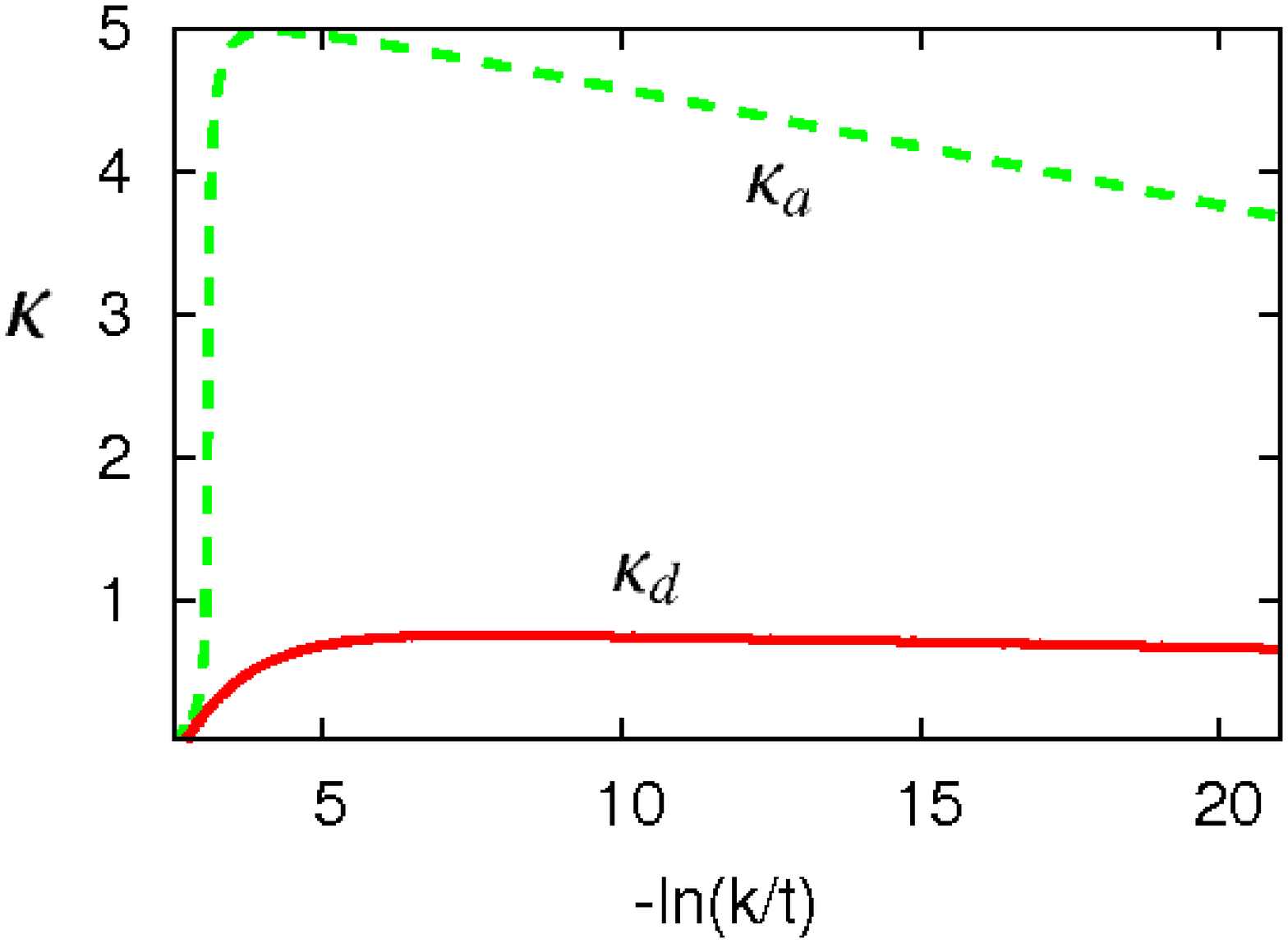}
\caption{\small{Upper panel: Flow of the renormalized antiferromagnetic order parameter $\kappa_a$ (green, dashed), renormalized $d$-wave superconducting order parameter $\kappa_d$ (red, solid) at $U/t=3$, $t'/t=-0.1$, $\mu/t=-0.72$ and $T/t=0.0131$. The dashed-dotted magenta curve shows the (unrenormalized) $d$-wave superconducting mass term $\bar m_d^2$ when it becomes nonzero again. Lower panel: Same as upper panel, but neglecting the mutual influence of the order parameters, i.\ e. each order parameter is set to zero in all contributions to the other boson as well as the inter-boson coupling $\lambda_{ad}$.}}
\label{order_parameters_1}
\end{figure}

The fact that both order parameters become zero during the flow at values of $k>k_{ph}$ in a temperature region around $T=0.0125t$ results from the mutual negative influence of the two types of order on each other. This influence is further illustrated by the upper and lower panels of Fig. \ref{order_parameters_1}, where the upper panel shows the flow of $\kappa_a$ together with that of $\kappa_d$ down to $k=k_{ph}$. For the temperature $T=0.0131t$ used in this graph $\kappa_a$ is nonzero during a much longer period of the renormalization flow than $\kappa_d$, which becomes zero at $-\ln(k/t)\approx4.6$ so that the superconducting mass term $\bar m_d^2$ (magenta curve) becomes nonzero again. The lower panel of Fig. \ref{order_parameters_1}, in contrast, shows the flow of the same couplings, but in this case the mutual influence of the order parameters has been neglected. This means that each order parameter is set to zero in all contributions to the other boson and the inter-bosonic quartic coupling $\lambda_{ad}$ is set to zero. As described in the previous section, this is equivalent to deriving the bosonic contributions to the flow equations from the $O(3)$- and $O(2)$-symmetric linear $\sigma$-models at finite temperature. According to this simplified treatment, neglecting the mutual influence of the two types of order, both order parameters remain nonzero down to $k=k_{ph}$.  Such a result would suggest a region of coexistence of ``global'' antiferromagnetic and $d$-wave superconducting order.

In our example this coexistence is destroyed by the mutual influence of the antiferromagnetic and superconducting bosons.  It is precisely this type of influence which has been taken into account in the upper panel of Fig. \ref{order_parameters_1}. We therefore conclude that the two types of order have a tendency to destroy each other. On the basis of a renormalized mean field treatment Ref. \onlinecite{reiss} (see in particular Figs. 10 and 11) reports on an analogous tendency of antiferromagnetism and superconductivity to mutually suppress each other.

\begin{figure}[t]
\includegraphics[width=65mm,angle=0.]{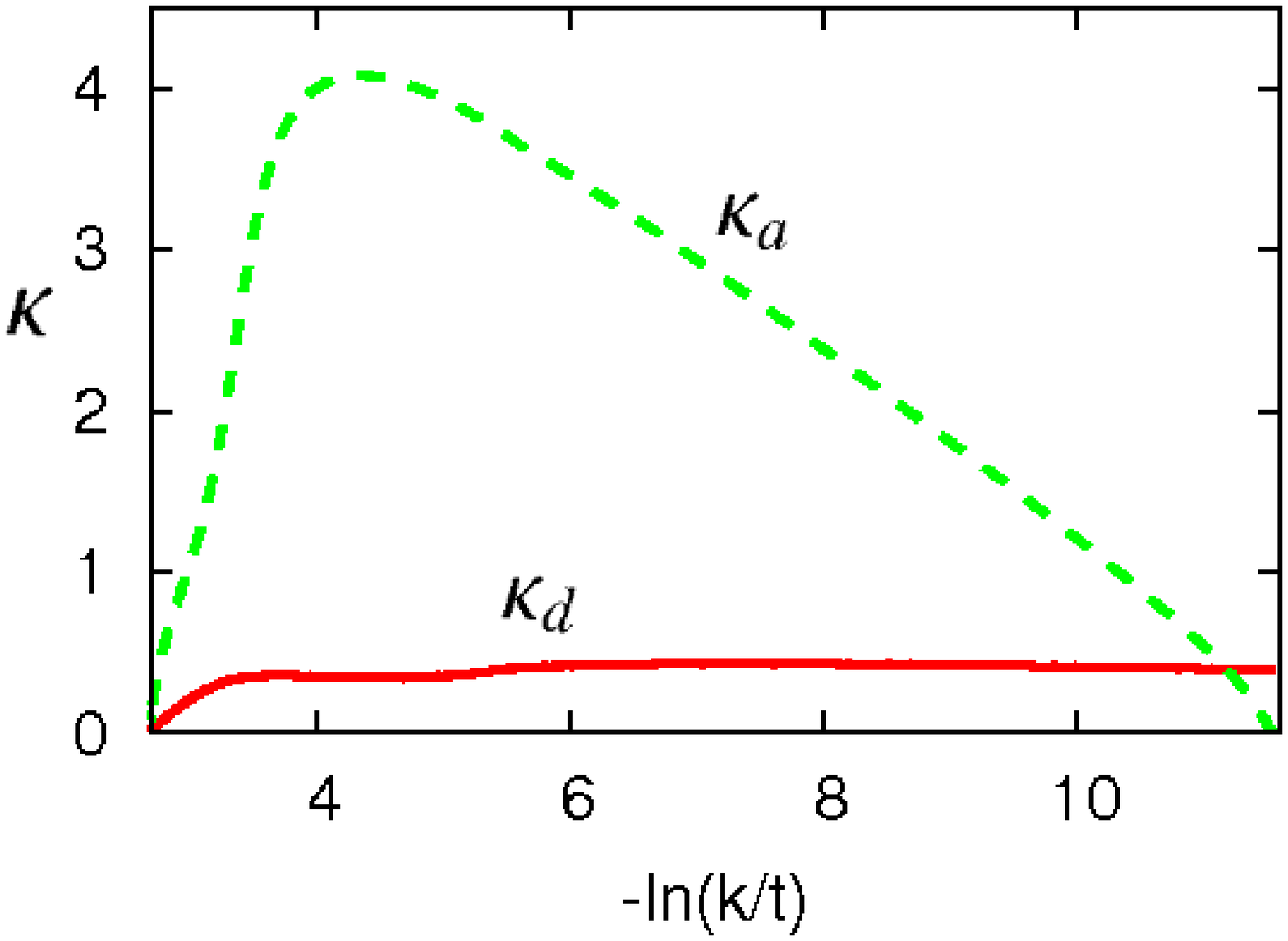}
\includegraphics[width=65mm,angle=0.]{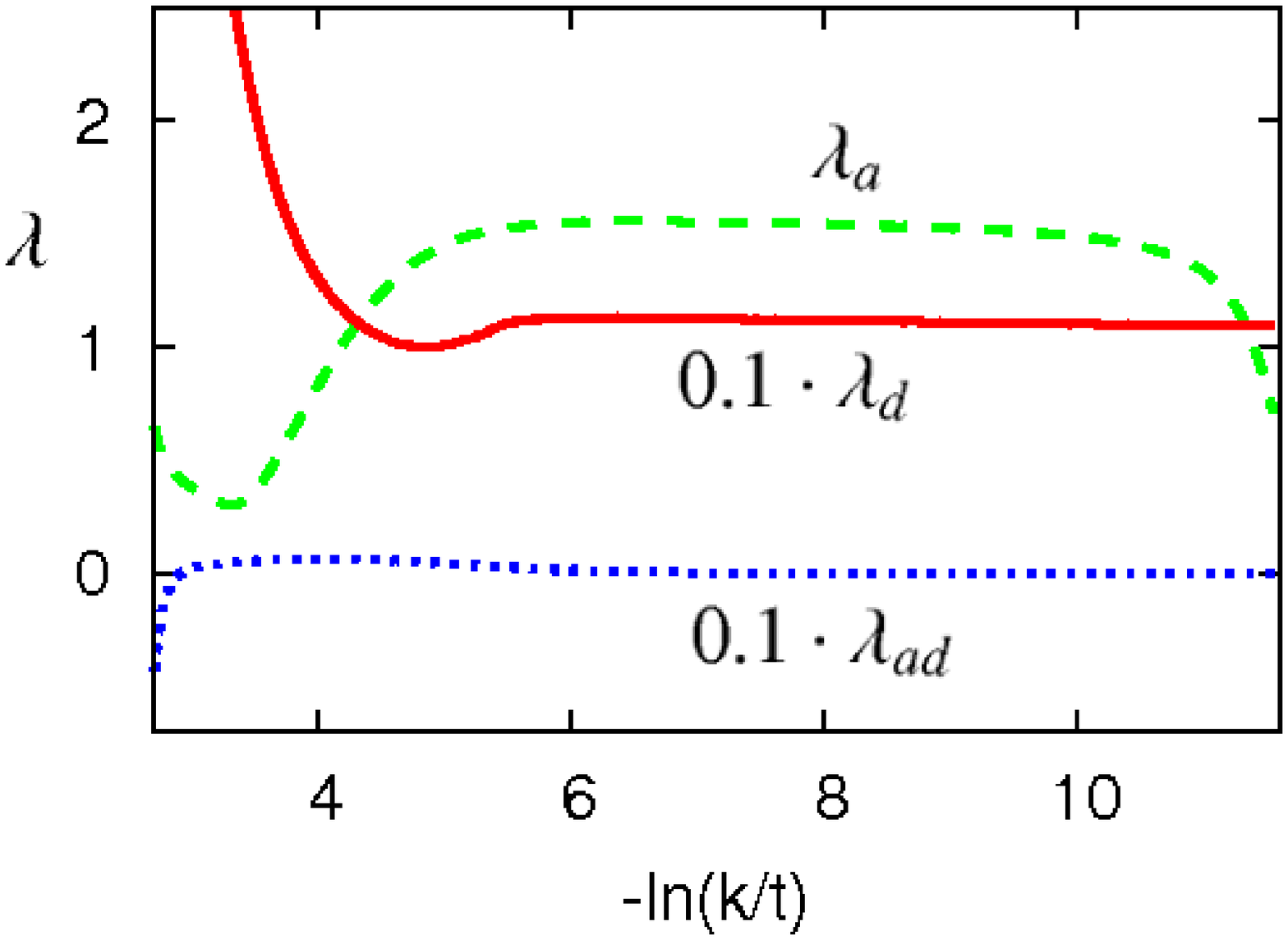}
\caption{\small{Upper panel: Flow of the renormalized antiferromagnetic order parameter $\kappa_a$ (green, dashed) and renormalized $d$-wave superconducting order parameter $\kappa_d$ (red, solid) at $U/t=3$, $t'/t=-0.1$, $\mu/t=-0.72$ and $T/t=0.0118$. Lower panel: Flow of the quartic couplings $\lambda_a$ (green, dashed), $\lambda_d$ (red, solid) and $\lambda_{ad}$ (blue, dotted), the latter two multiplied by a factor of $0.1$, for the same choice of parameters.}}
\label{order_parameters_2}
\end{figure}

For the curves shown in Fig. \ref{order_parameters_2} the temperature has been reduced in comparison to Fig. \ref{order_parameters_1}, so that both order parameters (upper panel) are nonzero for an important interval of the flow. Now $d$-wave superconducting order persists down to much lower scales $k$ of the renormalization flow. Although at an intermediate stage of the flow $\kappa_a$ is considerably larger than $\kappa_d$, it vanishes earlier during the flow due to the larger number of Goldstone modes for antiferromagnetism. In the lower panel of Fig. \ref{order_parameters_2} the flow of the quartic couplings $\lambda_a$, $\lambda_d$ and $\lambda_{ad}$ is displayed, where $\lambda_{ad}$ approaches zero much more quickly than $\lambda_a$ and $\lambda_d$ so that the two bosons are more or less independent and the flow is dominated by their Goldstone modes at low scales.

Taking things together, the example shown in the lower panel of Fig. \ref{orderparams} demonstrates that the phase transitions cannot always be understood by the universal behavior of linear or non-linear uncoupled $\sigma$-models. For example, in the $O(3)$-$\sigma$-model it is not possible to find a restoration of disorder at temperatures below the ones for which long-range order is realized. The competition of different bosons is crucial for a quantitative understanding of the phase diagram.

\subsection{Phase diagram}

We now turn to the discussion of the phase diagram obtained for $U=3t$ and $t'=-0.1t$, as shown in
Fig. \ref{phasediag}. For values of $-t'$ which are substantially smaller than $0.1t$ the quartic coupling $\lambda_a$ eventually becomes negative during the flow, which may indicate a tendency toward a first order transition which is not captured in the present truncation of the effective potential \eqref{Utrunc2}. For values of $-t'$ which are considerably larger, in contrast, the system exhibits a tendency toward ferromagnetism \cite{honisalmi}. In order to account for this instability, the truncation for the effective action and the parametrization of the bosonic propagators and Yukawa couplings specified in Appendix B of \cite{fourpoint} would have to be adjusted accordingly. Upon small variations of $t'$ the qualitative picture of the phase diagram remains essentially unchanged. If $-t'$ is reduced, all phase boundaries are shifted toward smaller values of $-\mu$, if $-t'$ is increased, they move into the other direction. For smaller values of the Hubbard interaction $U$, critical temperatures are lower and the phase boundaries are shifted in the direction of the van Hove filling chemical potential $\mu=4t'$. The results we have obtained for calculations at values of $U$ and $t'$ other than $U=3t$ and $t'=-0.1t$ do not alter the picture described in what follows.

At the van Hove filling we find a sizable difference between the pseudocritical temperature $T_{pc}$ and the true critical temperature $T_c$ for antiferromagnetism, which differ by a factor of about $2$, mainly due to the two antiferromagnetic Goldstone modes. In the $d$-wave superconducting regime at $-\mu/t>0.75$, in contrast, there is only a slight difference between $T_{pc}$ and $T_c$, in accordance with earlier results on the $O(2)$-symmetric model \cite{kw07}. The non-negligible difference between $T_{pc}$ and $T_c$ for $d$-wave superconductivity in the region between $-\mu/t=0.66$ and $-\mu/t=0.75$ is not due to Goldstone fluctuations but arises from the influence of antiferromagnetic order on the flow of $\kappa_d$.

One of the most intriguing questions about the phase diagram of the two-dimensional Hubbard model is whether it exhibits a region in parameter space where antiferromagnetic and $d$-wave superconducting order coexist. In principle the setup employed in the present work makes it possible to assess this question, but the results obtained by means of the truncation used here do not permit a definite answer. While they clearly suggest that the two types of order ``do not like each other'', they can hardly be taken to rule out the existence of a region in parameter space where antiferromagnetism and $d$-wave superconductivity coexist. Where the two phases border each other at low temperatures around $-\mu/t=0.66$ in Fig. \ref{phasediag} there is always at least one type of order which persists down to $k=k_{ph}=1\,\rm{cm}^{-1}\approx10^{-9}t$, and the value of $k$ where the second order parameter drops to zero is often very close to $k_{ph}$. Coexistence might occur, even within the truncation used here at temperatures below $T_{min}=4\cdot10^{-3}t$, the lowest temperature for which we have done calculations. In all cases where both order parameters remain finite for a considerable part of the flow, the values of the running couplings at $k=k_{ph}$ are highly sensitive to their values at the onset of the spontaneously broken regime. Therefore, we expect that further extensions of the truncation, which may influence the flow on intermediate scales, can have an important effect on the shape of the phase boundaries where the antiferromagnetic and $d$-wave superconducting phases are close to each other. Self-energy corrections, higher order bosonic couplings and the effect of the antiferromagnetic incommensurability, which we have neglected here in the SSB regimes, may be responsible for whether there exists a region in the phase diagram where antiferromagnetic and $d$-wave superconducting order coexist at $k=k_{ph}$. The results presented here, however, suggest that \textit{if} there is a region in $\mu\,-\,T$-space where antiferromagnetism and $d$-wave superconductivity coexist on a macroscopic level, this region is probably not very extended.

\section{Conclusions}
Our main conclusion is that functional renormalization combined with partial bosonization can give direct and physically transparent access to the low temperature behavior of the Hubbard model. We have computed the temperature dependence of the superfluid density or the antiferromagnetic order parameter for temperatures below the critical temperature from which the associated gaps for the electrons can easily be determined. Taking into account the fluctuations of composite bosons we can incorporate the important contributions from collective spin waves or $d$-wave superconducting bosons. This allows us to compute the critical temperature as a function of the chemical potential and therefore to establish the phase diagram.

The physics of collective boson fluctuations is quantitatively important. This is demonstrated in Fig. \ref{phasediag} by the substantial difference between the pseudocritical temperature (which is often associated with the critical temperature in other approaches) and the true critical temperature. Also the understanding of the region with coexisting local antiferromagnetic and superconducting order (in Fig. \ref{phasediag} for $-\mu/t$ between approximately $0.6$ and $0.8$) would be different without a proper understanding of the bosonic fluctuations.

It is obvious that more extended truncations can substantially improve the quantitative accuracy. One may include into the flowing momentum-dependent four-fermion vertex (instead of a constant $U$) the changes which cannot be absorbed by bosonization into the present channels, or one could increase the number of bosonic channels retained. Furthermore, a more accurate parametrization of the momentum and frequency dependence of the bosonic and fermionic propagators could be helpful. One may go beyond the quartic polynomial approximation for the effective bosonic potential, both in order to account for possible first order phase transitions and to give a more accurate description of the universal critical behavior for the second order phase transitions.

Nevertheless, we believe that several of our findings are rather robust. This concerns the existence of an extended region in the phase diagram for which local order but no long-range order exists. In particular, the transition between antiferromagnetic and superconducting order is rather complex, with coexisting local antiferromagnetic and superconducting order, but also a strong tendency of exclusion of coexisting long-range order due to the bosonic fluctuations.

\vspace{0.5cm}

{\bf Acknowledgments}: We would like to thank C. Husemann, K.-U. Giering, A. Katanin and M. Salmhofer for useful discussions. SF acknowledges support by Studienstiftung des Deutschen Volkes.

\renewcommand{\thesection}{}
\renewcommand{\thesubsection}{A{subsection}}
\renewcommand{\theequation}{A \arabic{equation}}


\end{document}